\documentclass[usenatbib, usegraphicx]{mn2e}

\usepackage{times, amssymb}
\usepackage[a4paper, totalwidth=480pt, totalheight=680pt]{geometry}

\newcommand{\vsini}{\mbox{\ensuremath{v\sin{i}}}}   
\newcommand{\kms}{\mbox{km s$^{-1}$}}   
\newcommand{\teff}{\mbox{$T_{\rm eff}$}}

\newcommand{\lama}{\mbox{$-79.0^{+4.5\tiny{\degr}} _{-4.3\tiny{\degr}}$}}
\newcommand{\lamtva}{\mbox{$-79.0$}}
\newcommand{\lamtea}{\mbox{$^{+4.5}_{-4.3}$}}

\newcommand{\vsinia}{\mbox{$5.77 \pm 0.35\, \kms$}}
\newcommand{\vsinitva}{\mbox{$5.77$}}
\newcommand{\vsinitea}{\mbox{$0.35$}}
\newcommand{\vsiniva}{\mbox{$5.79 $}}
\newcommand{\vsiniea}{\mbox{$0.35$}}

\newcommand{\rprstva}{\mbox{$0.10271$}}
\newcommand{\rprstea}{\mbox{$0.00058$}}
\newcommand{\rprsva}{\mbox{$0.10271$}}
\newcommand{\rprsea}{\mbox{$0.00059$}}

\newcommand{\arstva}{\mbox{$5.64$}}
\newcommand{\arstea}{\mbox{$0.13$}}
\newcommand{\arsva}{\mbox{$5.64 $}}
\newcommand{\arsea}{\mbox{$0.12$}}

\newcommand{\inctva}{\mbox{$88.65$}}
\newcommand{\inctea}{\mbox{$^{+0.51}_{-0.55}$}}
\newcommand{\incva}{\mbox{$88.07$}}
\newcommand{\incea}{\mbox{$1.15$}}

\newcommand{\periodtva}{\mbox{$2.5199449$}}
\newcommand{\periodtea}{\mbox{$0.0000013$}}
\newcommand{\periodva}{\mbox{$2.5199449 $}}
\newcommand{\periodea}{\mbox{$0.0000013$}}

\newcommand{\tzerotva}{\mbox{$54461.86099$}}
\newcommand{\tzerotea}{\mbox{$0.00024$}}
\newcommand{\tzerova}{\mbox{$54461.86099 $}}
\newcommand{\tzeroea}{\mbox{$0.00024$}}

\newcommand{\ktva}{\mbox{$0.1094$}}
\newcommand{\ktea}{\mbox{$0.00564$}}

\newcommand{\gammattva}{\mbox{$-13.2365$}}
\newcommand{\gammattea}{\mbox{$0.0027$}}

\newcommand{\gammaotva}{\mbox{$-13.5016$}}
\newcommand{\gammaotea}{\mbox{$0.0065$}}

\newcommand{\gammaootva}{\mbox{$-13.4990$}}
\newcommand{\gammaootea}{\mbox{$0.0057$}}

\newcommand{\lamb}{\mbox{$-4.7 \pm 4.0 \degr$}}
\newcommand{\lamtvb}{\mbox{$-4.7$}}
\newcommand{\lamteb}{\mbox{$4.0$}}

\newcommand{\vsinib}{\mbox{$7.32 \pm  0.88\, \kms$}}

\newcommand{\vsinitvb}{\mbox{$7.0$}}
\newcommand{\vsiniteb}{\mbox{$0.9$}}

\newcommand{\rprstvb}{\mbox{$0.10040$}}
\newcommand{\rprsteb}{\mbox{$0.00060$}}
\newcommand{\rprsvb}{\mbox{$0.10040$}}
\newcommand{\rprseb}{\mbox{$0.00060$}}

\newcommand{\arstvb}{\mbox{$5.98$}}
\newcommand{\arsteb}{\mbox{$0.15$}}
\newcommand{\arsvb}{\mbox{$5.90 $}}
\newcommand{\arseb}{\mbox{$0.33$}}

\newcommand{\inctvb}{\mbox{$83.64$}}
\newcommand{\incteb}{\mbox{$0.29$}}
\newcommand{\incvb}{\mbox{$83.64$}}
\newcommand{\inceb}{\mbox{$0.31$}}

\newcommand{\periodtvb}{\mbox{$2.3412127$}}
\newcommand{\periodteb}{\mbox{$0.0000020$}}
\newcommand{\periodvb}{\mbox{$2.3412124 $}}
\newcommand{\periodeb}{\mbox{$0.0000020$}}

\newcommand{\tzerotvb}{\mbox{$55081.37941$}}
\newcommand{\tzeroteb}{\mbox{$0.00017$}}
\newcommand{\tzerovb}{\mbox{$55081.37941 $}}
\newcommand{\tzeroeb}{\mbox{$0.00017$}}

\newcommand{\ktvb}{\mbox{$0.1482$}}
\newcommand{\kteb}{\mbox{$0.0025$}}

\newcommand{\gammattvb}{\mbox{$-17.7871$}}
\newcommand{\gammatteb}{\mbox{$0.0017$}}

\newcommand{\gammaotvb}{\mbox{$-17.8031$}}
\newcommand{\gammaoteb}{\mbox{$0.0040$}}

\newcommand{\gammaootvb}{\mbox{$-17.9050$}}
\newcommand{\gammaooteb}{\mbox{$0.0073$}}

\newcommand{\gammaoootvb}{\mbox{$-17.7905$}}
\newcommand{\gammaoooteb}{\mbox{$0.0019$}}

\newcommand{\lamc}{\mbox{$15^{+33\tiny{\degr}}_{-43\tiny{\degr}}$}}
\newcommand{\lamtvc}{\mbox{$15$}}
\newcommand{\lamtec}{\mbox{$^{+33}_{-43}$}}

\newcommand{\vsinic}{\mbox{$8.58 \pm 0.39\, \kms$}}
\newcommand{\vsinitvc}{\mbox{$8.58$}}
\newcommand{\vsinitec}{\mbox{$0.39$}}
\newcommand{\vsinivc}{\mbox{$8.69 $}}
\newcommand{\vsiniec}{\mbox{$0.40$}}

\newcommand{\rprstvc}{\mbox{$0.0844$}}
\newcommand{\rprstec}{\mbox{$0.0011$}}
\newcommand{\rprsvc}{\mbox{$0.0844$}}
\newcommand{\rprsec}{\mbox{$0.0011$}}

\newcommand{\arstvc}{\mbox{$12.15$}}
\newcommand{\arstec}{\mbox{$0.18$}}
\newcommand{\arsvc}{\mbox{$12.15$}}
\newcommand{\arsec}{\mbox{$0.19$}}

\newcommand{\inctvc}{\mbox{$88.83$}}

\newcommand{\incvc}{\mbox{$88.69$}}
\newcommand{\incec}{\mbox{$0.67$}}

\newcommand{\periodtvc}{\mbox{$ 6.871814$}}
\newcommand{\periodtec}{\mbox{$0.000045$}}
\newcommand{\periodvc}{\mbox{$ 6.871815$}}
\newcommand{\periodec}{\mbox{$0.000045$}}

\newcommand{\tzerotvc}{\mbox{$55335.92044$}}
\newcommand{\tzerotec}{\mbox{$0.00074$}}
\newcommand{\tzerovc}{\mbox{$55335.92050 $}}
\newcommand{\tzeroec}{\mbox{$0.00074$}}

\newcommand{\ktvc}{\mbox{$0.2538$}}
\newcommand{\ktec}{\mbox{$0.0035$}}

\newcommand{\gammattvc}{\mbox{$-9.8404$}}
\newcommand{\gammattec}{\mbox{$^{+0.0053}_{-0.0057}$}}

\newcommand{\gammaotvc}{\mbox{$-9.7181$}}
\newcommand{\gammaotec}{\mbox{$0.0063$}}

\newcommand{\gammaootvc}{\mbox{$-9.7951$}}
\newcommand{\gammaootec}{\mbox{$0.0027$}}

\newcommand{\lamd}{\mbox{$-9.7^{+9.0\tiny{\degr}}_{-7.7\tiny{\degr}}$}}
\newcommand{\lamtvd}{\mbox{$-9.7$}}
\newcommand{\lamted}{\mbox{$^{+9.0}_{-7.7}$}}

\newcommand{\vsinid}{\mbox{$11.8 \pm 0.5\, \kms$}}
\newcommand{\vsinitvd}{\mbox{$11.8$}}
\newcommand{\vsinited}{\mbox{$0.5$}}
\newcommand{\vsinivd}{\mbox{$11.5 $}}
\newcommand{\vsinied}{\mbox{$0.5$}}

\newcommand{\rprstvd}{\mbox{$0.09135$}}
\newcommand{\rprsted}{\mbox{$0.00089$}}
\newcommand{\rprsvd}{\mbox{$0.09110$}}
\newcommand{\rprsed}{\mbox{$0.00090$}}

\newcommand{\arstvd}{\mbox{$6.12$}}
\newcommand{\arsted}{\mbox{$^{+0.20}_{-0.21}$}}
\newcommand{\arsvd}{\mbox{$6.35 $}}
\newcommand{\arsed}{\mbox{$0.34$}}

\newcommand{\inctvd}{\mbox{$87.80$}}
\newcommand{\incted}{\mbox{$^{+0.75}_{-0.77}$}}
\newcommand{\incvd}{\mbox{$87.26$}}
\newcommand{\inced}{\mbox{$1.00$}}

\newcommand{\periodtvd}{\mbox{$3.0763370$}}
\newcommand{\periodted}{\mbox{$0.0000036$}}
\newcommand{\periodvd}{\mbox{$3.0763350 $}}
\newcommand{\perioded}{\mbox{$0.0000040$}}

\newcommand{\tzerotvd}{\mbox{$54437.67587$}}
\newcommand{\tzeroted}{\mbox{$ 0.00034$}}
\newcommand{\tzerovd}{\mbox{$54437.67582 $}}
\newcommand{\tzeroed}{\mbox{$0.00034$}}

\newcommand{\ktvd}{\mbox{$0.1580$}}
\newcommand{\kted}{\mbox{$0.0041$}}

\newcommand{\gammattvd}{\mbox{$-22.3650$}}
\newcommand{\gammatted}{\mbox{$0.0040$}}

\newcommand{\gammaotvd}{\mbox{$-0.0937$}}
\newcommand{\gammaoted}{\mbox{$0.0031$}}

\title[The spin-orbit alignment of WASP-1b, WASP-24b, WASP-38b and HAT-P-8b]
{The spin-orbit angles of the transiting exoplanets WASP-1b, WASP-24b,  WASP-38b and HAT-P-8b from Rossiter-McLaughlin observations}

\author[E. K. Simpson et al.]
{E. K.~Simpson$^1$\thanks{Email: esimpson05@qub.ac.uk.}\thanks{This work is based on observations collected with the SOPHIE spectrograph on the 1.93 m telescope at Observatoire de Haute-Provence (CNRS), France, by the SOPHIE Consortium;  the Nordic Optical Telescope, operated on the island of La Palma jointly by Denmark, Finland, Iceland, Norway, and Sweden, in the Spanish Observatorio del Roque de los Muchachos of the Instituto de Astrofisica de Canarias; and the HARPS spectrograph mounted on the ESO 3.6m at the La Silla Observatory in Chile under proposal 084.C-0185. }, D.~Pollacco$^1$, A.~Collier Cameron$^2$, G.~H\'{e}brard$^{3,4}$, D. R. Anderson$^5$,  
\newauthor  S. C. C.~Barros$^1$, I.~Boisse$^3$, F.~Bouchy$^{3,4}$, F.~Faedi$^{1}$, M. Gillon$^6$, L.~Hebb$^{7}$, F. P.~Keenan$^1$,
\newauthor   G. R. M.~Miller$^2$, C.~Moutou$^{8}$, D. Queloz$^{9}$, I.~Skillen$^{10}$, P.~Sorensen$^{11}$, H. C.~Stempels$^{12}$, 
\newauthor  A.~Triaud$^{9}$, C. A.~Watson$^1$, P. A.~Wilson$^{11,13}$ \\
$^1$ Astrophysics Research Centre, School of Mathematics \& Physics, QueenÕs University Belfast, BT7 1NN, UK\\
$^2$ School of Physics and Astronomy, University of St Andrews, North Haugh, St Andrews, Fife KY16 9SS, UK\\
$^3$ Institut d'Astrophysique de Paris, UMR7095 CNRS, Universit\'e Pierre \& Marie Curie, France\\
$^4$ Observatoire de Haute-Provence, CNRS/OAMP, 04870 St Michel l'Observatoire, France\\
$^5$ Astrophysics Group, Keele University, Staffordshire, ST5 5BG, UK\\
$^6$ Universit\'e de Li\`ege, All\'ee du 6 ao\^ut 17, Sart Tilman, Li\`ege 1, Belgium\\
$^7$ Department of Physics and Astronomy, Vanderbilt University, Nashville, TN 37235, USA\\
$^8$ Laboratoire d'Astrophysique de Marseille, 38 rue Fr\'ed\'eric Joliot-Curie, 13388 Marseille cedex 13, France\\
$^{9}$ Observatoire de Gen\`eve, Universit\'e de Gen\`eve, 51 Ch. des Maillettes, 1290 Sauverny, Switzerland\\
$^{10}$ Isaac Newton Group of Telescopes, Apartado de Correos 321, E-38700 Santa Cruz de la Palma, Spain \\
$^{11}$ Nordic Optical Telescope, Apartado de Correos 474, E-387 00 Santa Cruz de la Palma, Canary Islands, Spain\\
$^{12}$ Department of Physics and Astronomy, Uppsala University, Box 516, SE-751 20 Uppsala, Sweden\\
$^{13}$ Astrophysics Group, School of Physics, University of Exeter, Stocker Road, Exeter, EX4 4QL}

\begin{document}

\maketitle

\begin{abstract}
We present observations of the Rossiter-McLaughlin effect for the transiting exoplanet systems WASP-1, WASP-24, WASP-38 and HAT-P-8, and deduce the orientations of the planetary orbits with respect to the host stars' rotation axes. The planets WASP-24b, WASP-38b and HAT-P-8b appear to move in prograde orbits and be well aligned, having sky-projected spin orbit angles consistent with zero: $\lambda$ = \lamb, $\lambda$ = \lamc\ and $\lambda$ = \lamd, respectively. The host stars have \teff\ $<$ 6250 K and conform with the trend of cooler stars having low obliquities. WASP-38b is a massive planet on a moderately long period, eccentric orbit so may be expected to have a misaligned orbit given the high obliquities measured in similar systems. However, we find no evidence for a large spin-orbit angle. By contrast, WASP-1b joins the growing number of misaligned systems and has an almost polar orbit, $\lambda$ =  \lama . It is neither very massive, eccentric nor orbiting a hot host star, and therefore does not share the properties of many other misaligned systems.

\end{abstract}

\begin{keywords}
stars: planetary systems -- stars: individual: WASP-1,  WASP-24, WASP-38, HAT-P-8 -- techniques: radial velocities
\end{keywords}

\section{Introduction}\label{Intro}

The process of exoplanet migration has been a hotly debated topic since the first close-in planets were discovered \citep{MQ95,BM96,MB96}. We are able to explore the mechanisms which move planets inwards from large orbits through a statistical analysis of their dynamical properties such as orbital period, eccentricity and spin-orbit alignment. In particular, transiting planets allow us to measure the sky-projected angle ($\lambda$) between the stellar rotation axis and planetary orbit through measurement of the Rossiter-McLaughlin (RM) effect  \citep{Rossiter24,McLaughlin24}. The effect is caused by the planet sequentially passing over and blocking portions of the rotating stellar surface resulting in a radial velocity (RV) shift which traces the trajectory of the planet across the stellar disc and allows the orbital obliquity to be estimated. 

The first measurements of $\lambda$ found the systems to be well aligned and suggested that the planets had lost orbital angular momentum through interactions with the proto-planetary disc \citep{Lin96,Murray98}. However, there are now a growing number of planets with highly misaligned and even retrograde orbits which now make up approximately one-third of the systems so far studied. Several theories have been postulated to explain this. For example, it has been suggested that proto-planetary discs may not always be aligned with the stellar rotation axis as previously assumed \citep*{Bate10, Lai10}, although \citet{Watson10disk} found no evidence for misaligned debris-discs in eight systems. \citet*{FW09} and \citet{Triaud10} suggest that another, more dynamically violent process involving interactions with a third body (another planet or star), causes the misaligned orbits \citep[see][]{RF96,WM03,Nagasawa08}. Whether a single mechanism or a combination of several is at work remains to be tested by increasing the number of measured systems.

Misaligned orbits appear to be synonymous with eccentricity; six of the eight eccentric systems with measured spin-orbit angles are reported to have obliquities significantly different from zero, XO-3 \citep{Hebrard08}, HD 80606 \citep{Moutou09, Pont09b, Gillon09, Winn09b}, WASP-8 \citep{Queloz10}, WASP-14 \citep{Johnson09}, HAT-P-11 \citep{Winn10hatp11}, HAT-P-14 \citep{Winn10hatp14} with HAT-P-2 \citep{Loeillet08} and HD 17156 \citep{Narita09hd17156} being the exceptions. Although HD 17156 does not show a large sky-projected misalignment, \citet{Sman10} notes that the stellar rotation axis may be tilted along the line-of-sight. Another trend, noted by \citet{Johnson09}, is the correlation between planet mass and misalignment. \citet{Hebrard10} suggests that there could be several populations of planets, with those more massive than Jupiter undergoing a different migration scenario leading to the high spin-orbit angles. \citet{Winn10} and \citet{Sman10} find that misaligned orbits are more common in host stars with larger masses/higher effective temperatures. It is suggested that the tidal torques experienced by cooler stars, with deeper convective zones, could cause their envelopes to quickly align with a planet's orbit, thereby erasing any initial misalignment. Exceptions to this effect may be longer period and low mass planets which experience weaker tidal forces and therefore longer tidal timescales.  

We present spectroscopic observations of the transiting planets WASP-1b \citep{Cameron07}, WASP-24b \citep{Street10}, WASP-38b \citep{Barros10} and HAT-P-8b \citep{Latham09} obtained using the HARPS, SOPHIE and FIES spectrographs to determine their spin-orbit alignments. All the host stars are of similar temperature, $\sim$ 6100 K, but have diverse physical and dynamical properties which allows us investigate possible trends independent of \teff. WASP-1, WASP-24 and HAT-P-8 are non-eccentric, short period planets (2.3--3.1 d) with masses similar to Jupiter (0.9--1.5 M$_{\rm J}$), whereas WASP-38 is a massive planet on a moderately longer period orbit with a small but significant eccentricity ($m_{\rm p}$ = 2.7 M$_{\rm J}$, $P$ = 6.9 d, $e$ = $0.032^{+0.0050}_{-0.0044}$, \citealt{Barros10}). 

In Section \ref{OandM} we describe the general data analysis procedures and methods performed to measure the spin-orbit angles. The derived parameters for the four systems are presented in Section \ref{Analysis}, and the implications of these are discussed in Section \ref{Discussion}.

\section{Data Analysis} \label{OandM}
 
\subsection{Radial velocity extraction}

The data presented in this paper were obtained using three spectrographs; HARPS, SOPHIE and FIES, the instrumental setup and data reduction for each is described in this section. Details of the individual observations and results of each system are discussed in Section \ref{Analysis}.  

\subsubsection{HARPS}

The HARPS instrument is a high resolution (R = 110,000) stabailed echelle spectrograph mounted at the La Silla 3.6 m ESO telescope. Observations were conducted in the OBJO mode without simultaneous Thorium-Argon (ThAr) calibration. The wavelength solution was calculated using a ThAr calibration at the start of the night, and HARPS is stable within 1 m s$^{-1}$ across a night \citep{R04}, which is much lower than the photon noise on the data points. Spectra were extracted and cross-correlated against a template of a G2V-type star using the HARPS Data Reduction Software (DRS), see \citet{Baranne96}, \citet{Pepe02}, \citet{Mayor03} and \citet{LovisPepe07} for more details.  

\subsubsection{SOPHIE}

SOPHIE is a cross-dispersed, environmentally-stabilised echelle spectrograph (wavelength range 3872.4--6943.5\AA) designed for  high-precision radial velocity measurements. The spectrograph was used in high efficiency mode (resolution $R$ = 40,000) with the CCD in slow read-out mode to reduce the read-out noise. Two 3 arc-second diameter optical fibres were used, the first centred on the target and the second on the sky to simultaneously measure its background in case of contamination from scattered moonlight. The spectra were reduced using the SOPHIE pipeline \citep{Perruchot08}. Radial velocities were computed from a weighted cross correlation of each spectrum with a numerical mask of G2V spectral type.  A Gaussian was fitted to the cross-correlation functions to obtain the radial velocity shift, and the uncertainty was computed using the empirical relation given by \citet{Bouchy09} and \citet{Cameron07}.

\subsubsection{FIES}

The FIES spectrograph is mounted on the 2.5-m Nordic Optical Telescope on La Palma. FIES was used in medium resolution mode ($R$ = 46,000) with simultaneous ThAr calibration.  Spectra were extracted using the bespoke data reduction package FIEStool\footnote{http://www.not.iac.es/instruments/fies/fiestool/FIEStool.html}.  An IDL cross-correlation routine was used to obtain the radial velocities by finding the maximum of the cross-correlation functions of 30 spectral orders and taking the mean. 

A template spectrum was constructed by shifting and co-adding the out-of-transit spectra, against which the individual spectra were cross-correlated to obtain the final velocities. This template was cross-correlated with a high signal-to-noise spectrum of the Sun to obtain the absolute velocity to which the relative RVs were shifted. We estimated the RV uncertainty by $\sigma$ = RMS$(v)$ /  $\sqrt{N}$,  where $v$ is the RV of the individual orders and $N$ is the number of orders.

When fitting the data, we found that the FIES observations has a reduced $\chi^{2}_{red}$ = $\chi^{2}$/dof $>$ 1 (dof = number of points -- number of fitted parameters). This suggests that there is an extra source of noise present which has not been accounted for in the internal errors. The cause may be that the ThAr calibration does not travel through the same light path as the stellar light \citep{Buchhave10}, or that because the fibre lacks a scrambler there is non-uniform illumination of the spectrograph \citep{Queloz99}. To account for these instrumental effects, 15 m s$^{-1}$ of uncorrelated noise needed to be added in quadrature to the internal error estimation to obtain a reduced $\chi^{2}$ value of unity. The photon errors are given in the data tables. whereas the rescaled uncertainties are shown in the figures. 

\subsection{Model fitting}

\begin{figure}
\begin{center}
\includegraphics[angle=0, width=\columnwidth]{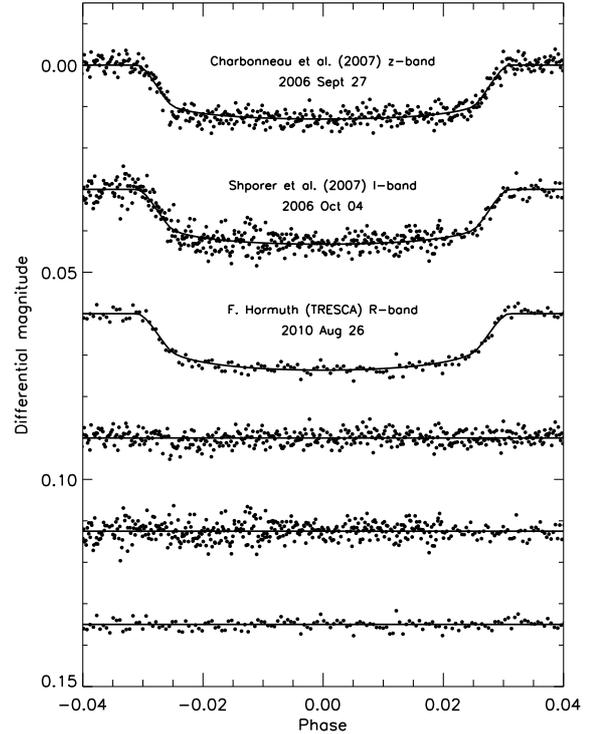}
\caption{Top: Photometry of three transits of WASP-1b and the residuals from the best-fit model.}
\label{WASP1phot}
\end{center}
\end{figure}

The RM effect and orbit were fitted simultaneously using all the available spectroscopic data. A Keplerian model was used for the orbit, and the analytical approach described in \citet*{Ohta05} was used to calculate the RM effect. We refer the reader to this paper for a detailed description of the model. The equations were modified to make them dependent on $R_{p}/R_{*}$ and $a/R_{*}$ rather than $R_{p}$, $R_{*}$ and $a$, to reflect the parameters derived from photometry and reduce the number of free parameters. 

We modelled the RM velocity anomaly as $v_{\rm RM}  =  - \delta v_{p}/(1-\delta)$ where $\delta$ is flux blocked by planet $\sim(R_{p}/R_{*})^{2}$ and $v_{p}$ is sub-planet velocity, i.e. the velocity component of the rotating stellar surface blocked by the planet $\sim x_{p} ~ \vsini ~ /R_{*} $, where $ x_{p}$ is the x co-ordinate of the position of the planet on the stellar surface, see Figure 5 of \citet*{Ohta05}.

In brief, the model comprises the following  parameters: the orbital period, $P$; mid-transit time (in the UTC system), $T_\mathrm{0}$; planetary to stellar radius ratio, $R_{p}/R_{*}$; scaled semi-major axis $a/R_*$;  orbital inclination, $i$; orbital eccentricity, $e$; longitude of periastron, $\omega$; radial velocity semi-amplitude, $K$; systemic velocity of orbital dataset, $\gamma$; sky projected angle between the stellar rotation axis and orbital angular momentum vector, $\lambda$; projected stellar rotational velocity, $v \sin {i} $; and the stellar linear limb-darkening coefficient, $u$. 

Each dataset was allowed to have a different systemic velocity ($\gamma$) to account for instrumental offsets. A linear limb darkening law was assumed, as the quadratic law alters the model by only a few m s$^{-1}$ and so does not seem justified given the precision of the RM data. As a test, $u$ was left as a free parameter and no significant effect on $\lambda$ or $\vsini$ was found. 

Some parameters have been tightly constrained by previous observations (e.g. $P$, $R_{\rm p}$/$R_{*}$). We use this information in the form of a penalty function on the $\chi^{2}$ statistic: 

\begin{eqnarray}
&  \chi^2   & =  \sum_{i}^{} \left [ \frac{v_{i, \mathrm{obs}} - v_{i, \mathrm{calc}}}{\sigma_{i}} \right ]^{2}  + \;\;\;\nonumber \\
    &&\left(  \frac{A - A_{\rm obs}+ \sigma_{Aobs} \times G(0,1) }{\sigma_{\rm Aobs}} \right)  ^{2}
\end{eqnarray}

\noindent where $v_{i, \mathrm{obs}}$ and $v_{i, \mathrm{calc}}$ are the $i$th observed and calculated radial velocities and $\sigma_{i}$ is the corresponding observational error. $A$ is a fitted parameter, $A_{\rm obs}$ is the parameter value determined from other observations and  $\sigma _{\rm Aobs}$ is the uncertainty in $A_{\rm obs}$.  The value $G(0,1)$ is a Gaussian randomly generated number of mean = 0 and standard deviation = 1. This allows the uncertainty in $A_{\rm obs}$ to be accounted for in the error budget. Depending on the situation, several penalty functions were used to constrain parameters in the fit and these are described in the individual analyses. If a parameter has asymmetric uncertainties, we have adopted the larger error.    

Best-fit parameters were obtained by minimising the $\chi^2$ statistic using the IDL-based MPFIT function \citep{mpfit}; a least-squares minimisation technique using the Levenberg-Marquardt algorithm. The $1\sigma$ best-fit parameter uncertainties were calculated using a Monte-Carlo method.  We created $10^{5}$ synthetic data sets by adding a 1$\sigma$ Gaussian random variable to the data points. The free parameters were re-optimised for each simulated data-set to obtain the distribution of the parameter values. These distributions were not assumed to be Gaussian and the 1$\sigma$ limits were found from the 15.85\% and 84.15\% bounds.

\begin{figure*}
\begin{center}
\includegraphics[angle=0, width=\columnwidth]{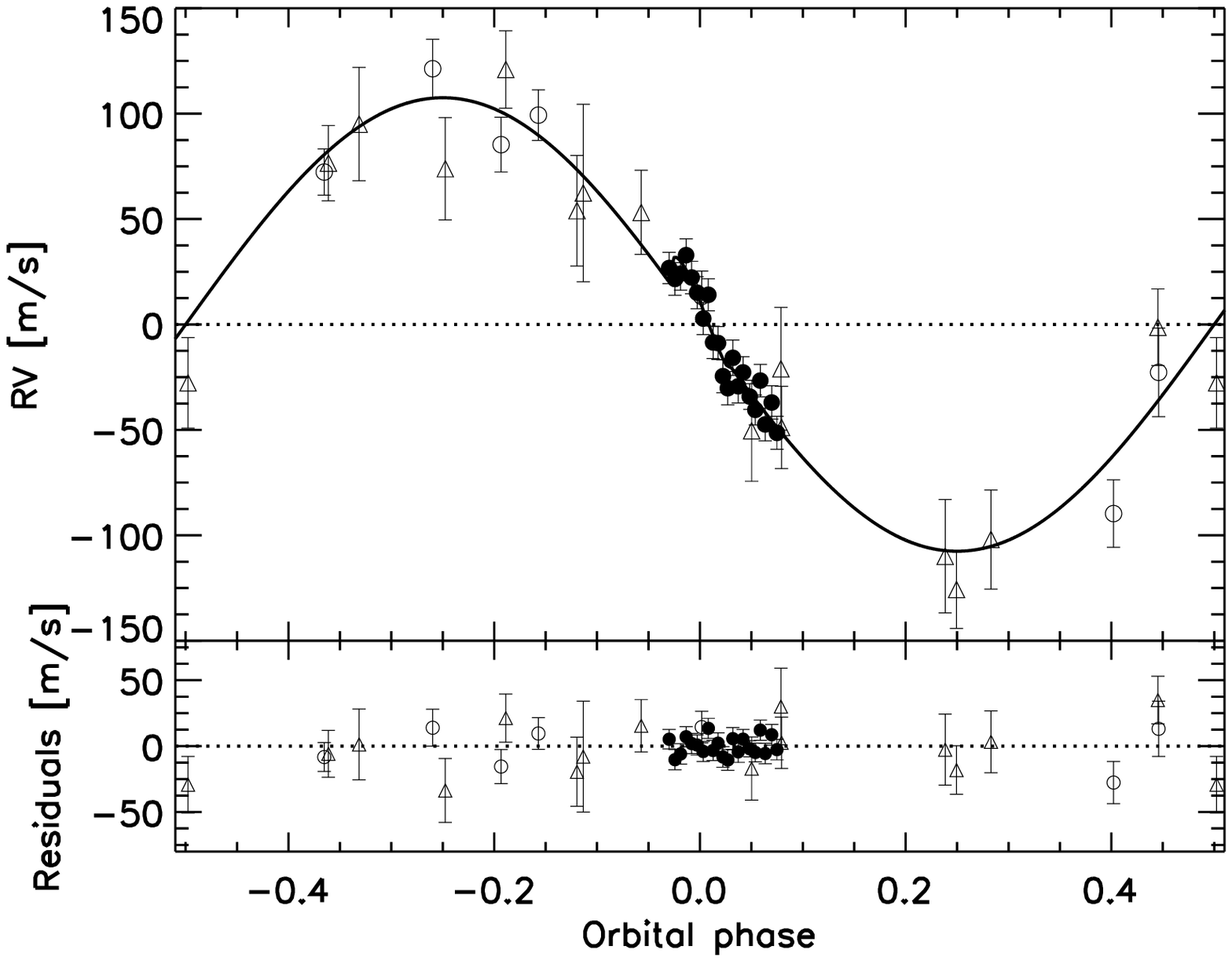}
\includegraphics[angle=0, width=\columnwidth]{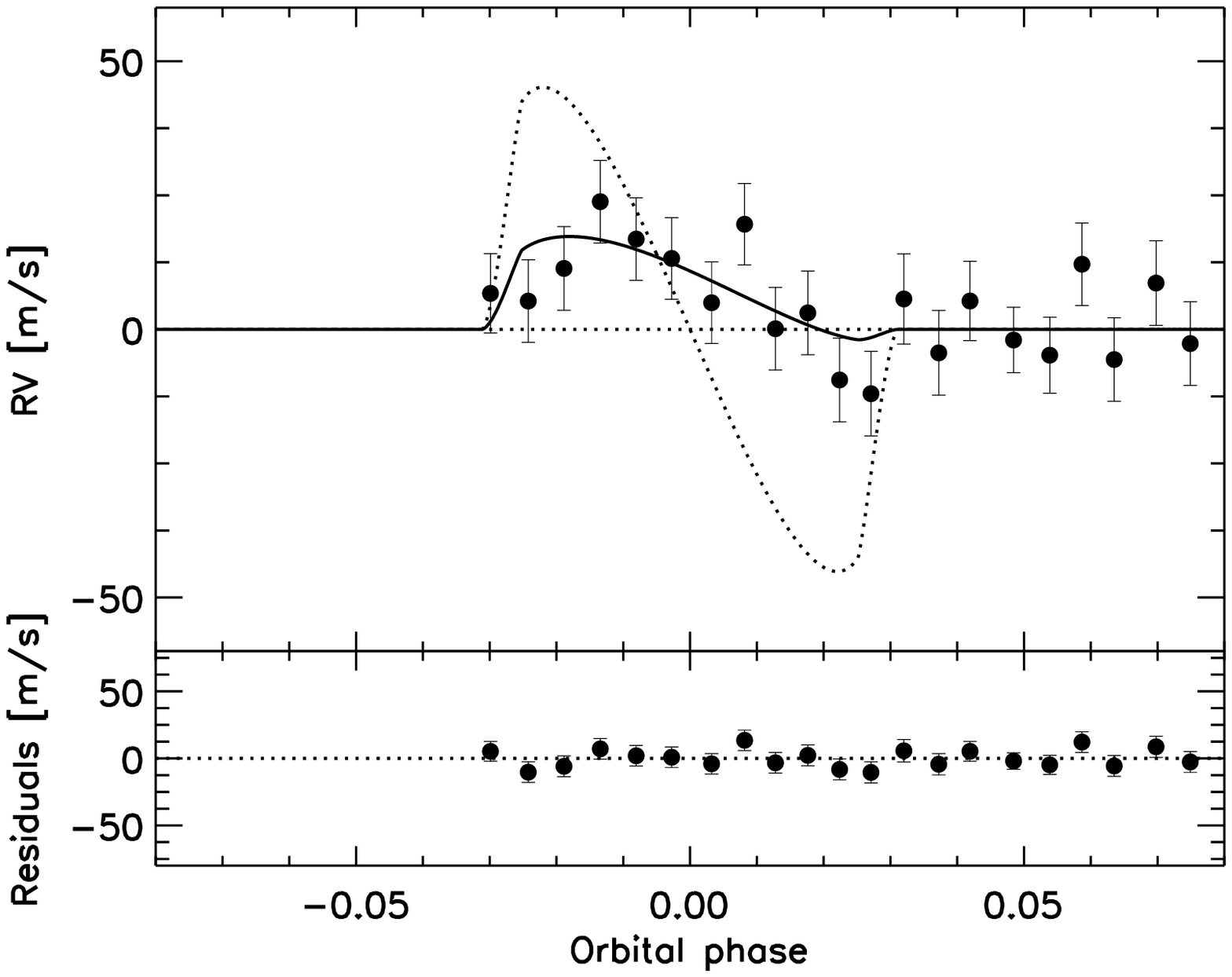}
\vspace{0.3cm}
\caption{Left: Phase folded radial velocities of WASP-1 minus the systematic velocity (given in Table \ref{Resultsw1}), over-plotted with the best-fit model, with the residuals shown below. The orbital observations were taken using FIES (open triangles) and SOPHIE (open circles) and the transit sequence using SOPHIE (filled circles). Right: Spectroscopic transit minus the orbital velocity, over-plotted with the best-fit model ($\lambda = \lama, \vsini = \vsinia$, and residuals shown below. The dotted line represents the RM effect of an aligned orbit. }
\label{WASP1figure}
\end{center}
\end{figure*}

\section{Analysis}\label{Analysis}

\subsection{WASP-1}\label{WASP1}

A transit of WASP-1b was observed with the SOPHIE spectrograph at the 1.93m telescope at Haute-Provence Observatory on the night of 2009 September 24. We acquired 21 spectra of WASP-1 covering the full transit (224 minutes, 12 observations) and a period of duration 156 minutes (9 observations) post transit. No observations were possible prior to transit due the the low altitude of the target at this time. The exposure time was adjusted between 13 and 22 minutes in order to reach a constant signal-to-noise ratio (S/N) of 30 at 550 nm. Atmospheric conditions were stable and the exposure time did not require altering to obtain the required S/N. Moon illumination was 39\% and at a distance of $>115\degr$ so this did not significantly affect the radial velocity determination. 

To fit the orbit, we used 7 SOPHIE observations obtained during the discovery of the planet \citep{Cameron07} and 16 FIES spectra taken at various orbital phases between 2009 January 1 and 2009 September 27. The new SOPHIE and FIES observations are shown in Table \ref{RVsW1}. \citet{Wheatley10} observed a secondary eclipse of WASP-1 and found that the eccentricity was consistent with zero, and in the models, $e$ has been set to 0 accordingly. The linear limb-darkening coefficient was chosen from the tables of \citet{Claret04} (ATLAS models) for the $g'$ filter and fixed at $u = 0.73$. 

A significant period of time has passed since the discovery of WASP-1 in 2006 and the orbital ephemeris may have drifted, leading to an inaccurate determination of the mid-transit time. To update the ephemeris, we fitted the z-band light curve presented in \citet{Charbonneau07}, the full I-band transit from \citet{Shporer07} and a high quality light curve taken on 2010 August 26 by F. Harmuth as part of the TRESCA/ETD project\footnote{http://var2.astro.cz/EN/tresca/} \citep*{Poddany10}, as shown in Figure \ref{WASP1phot}. The TRESCA transit was observed using the 1.2-m telescope at Calar Alto observatory, Spain using an R-band filter. A Markov Chain Monte Carlo (MCMC) routine was used to fit the data, see \citet{Cameron07} and \citet{Pollacco08} for more details. The non-linear limb darkening coefficients were chosen for the appropriate stellar temperature and photometric passband for each light curve. The parameters found for $P$, $T_{\rm 0}$,  $R_{\rm p}/R_{*}$, $a/R_{*}$ and $i_{\rm p}$ were used to constrain the fit in the form of penalty functions, and are shown in Equation \ref{eqw1}. 

WASP-1b has a low impact parameter, $b < $ 0.2, and in this regime $\vsini$ and $\lambda$ are highly correlated. The shape of the RM signal is not strongly dependent on $\lambda$, whereas the amplitude is dependent on both \vsini\ and $\lambda$. To break this degeneracy we introduced a penalty function on \vsini\ using the value of 5.79 $\pm$ 0.35 \kms\ found from spectroscopic line broadening measurements by \citet{Stempels07}. The following $\chi^{2}$ statistic was adopted:

\begin{eqnarray} \label{eqw1}
& \chi^2 & =  \sum_{i}^{} \left [ \frac{v_{i, \mathrm{obs}} - v_{i, \mathrm{calc}}}{\sigma_{i}} \right ]^{2}  + \;\;\;\nonumber \\
&  &    \left(  \frac{v\sin{i} - \vsiniva\ \kms +  \vsiniea ~ \kms  \times G(0,1)}{\vsiniea~ \kms} \right)  ^{2} + \;\;\;\nonumber \\
&  &     \left(  \frac{P - \periodva ~{\rm d}  + \periodea~{\rm d} \times G(0,1) }{\periodea ~{\rm d}} \right)  ^{2}   + \;\;\;\nonumber \\
&  &    \left(  \frac{T_{\rm0} - \tzerova + \tzeroea \times  G(0,1)}{\tzeroea} \right)  ^{2} +   \;\;\;\nonumber \\
&  &   \left(  \frac{R_{\rm p}/R_{*} - \rprsva + \rprsea \times G(0,1) }{\rprsea} \right)  ^{2} + \;\;\;\nonumber \\ 
&  &      \left(  \frac{a/R_{*} - \arsva  + \arsea \times G(0,1)}{\arsea} \right)  ^{2} + \;\;\;\nonumber \\
&  &    \left(  \frac{i_{\rm p} -  \incva \degr +  \incea \degr \times G(0,1)}{\incea \degr} \right)^{2} 
\end{eqnarray}

\begin{figure*}
\begin{center}
\includegraphics[angle=0, width=\columnwidth]{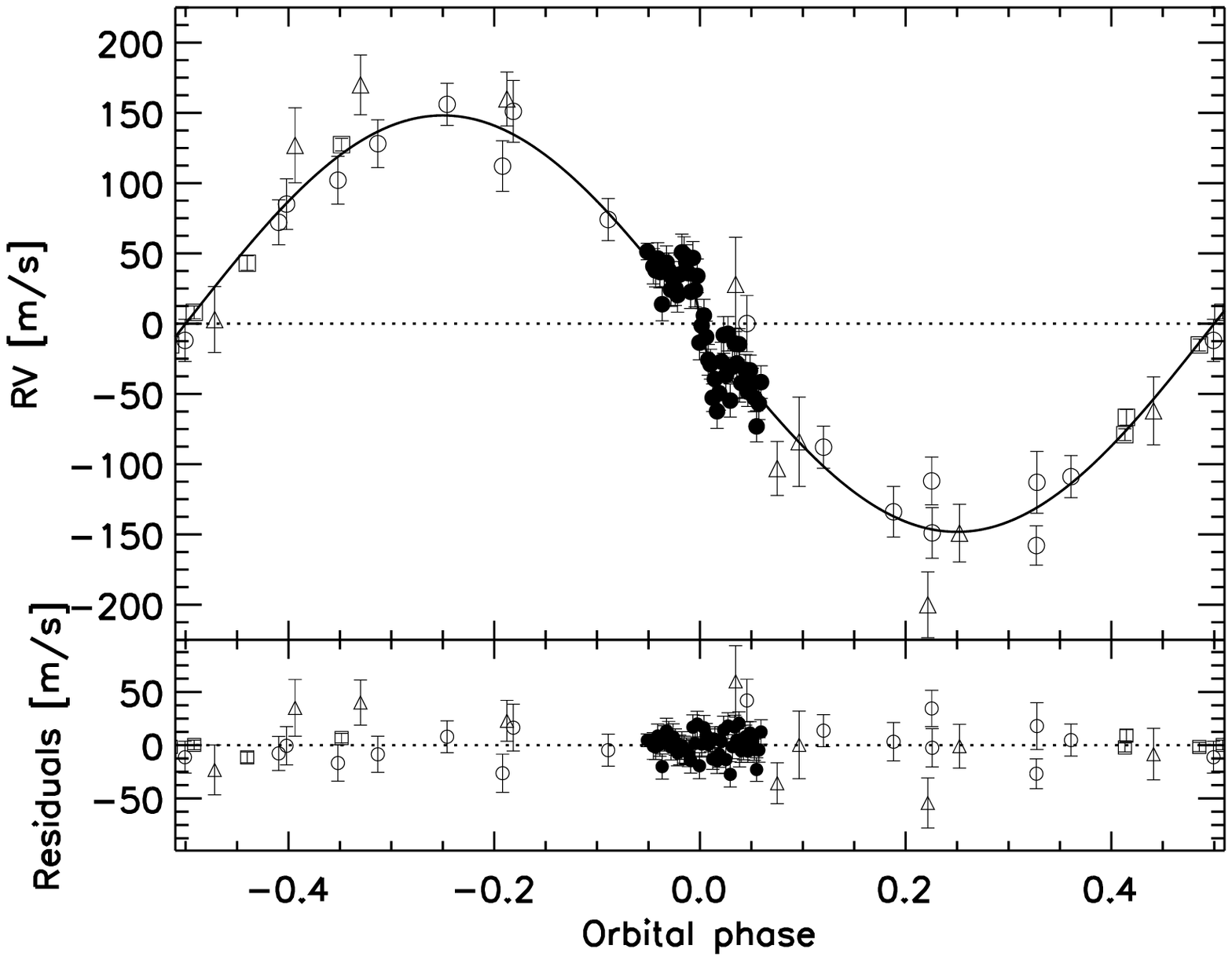}
\includegraphics[angle=0, width=\columnwidth]{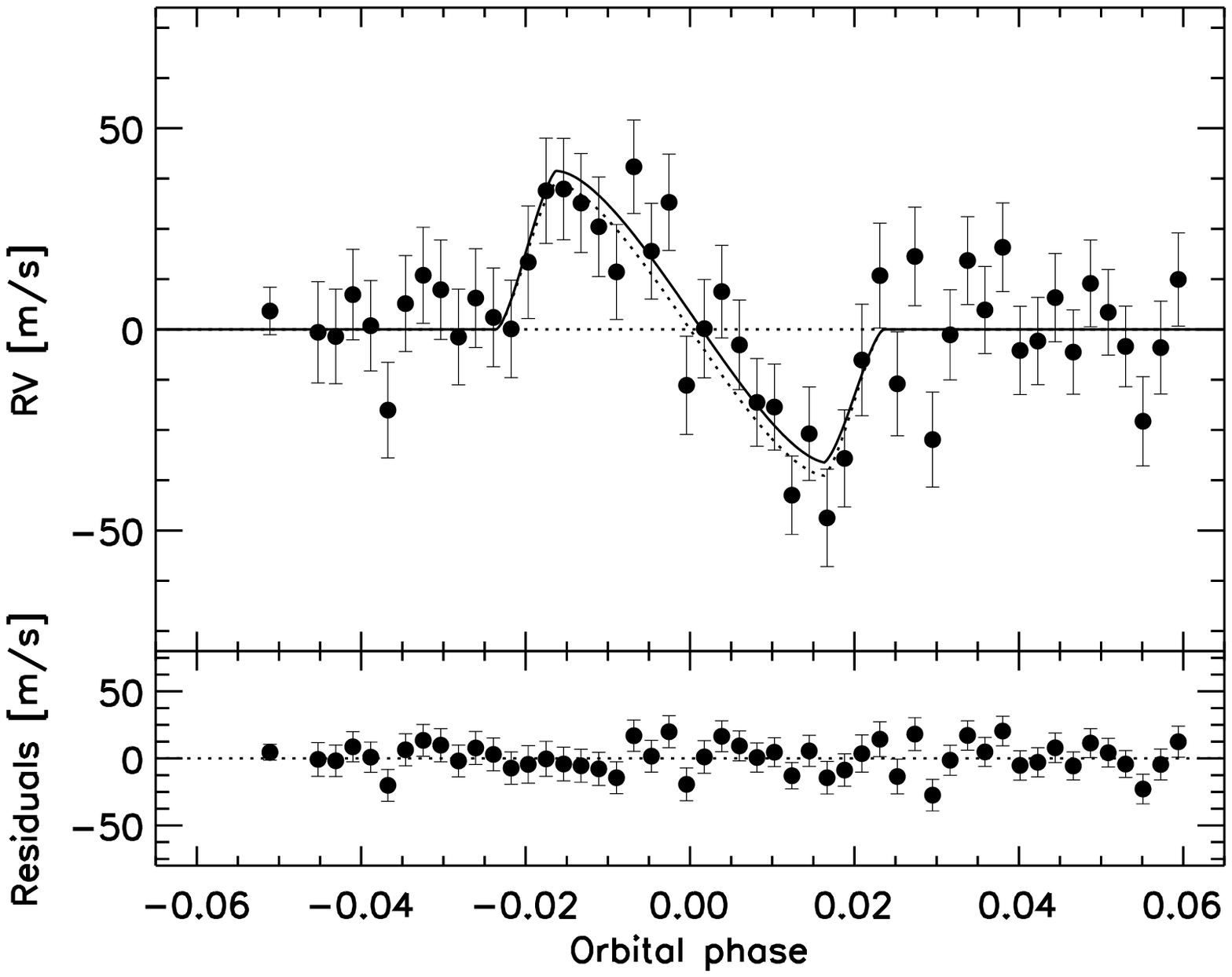}
\vspace{0.3cm}
\caption{Left: Phase-folded radial velocities of WASP-24 over-plotted with the best-fit model. The systematic velocity has been removed (see Table \ref{Resultsw24}) and the residuals are shown below. The orbital points were taken with FIES (open triangles), CORALE (open squares) and HARPS (open squares) and the HARPS transit sequence is shown in filled circles. Right: The RM effect minus the orbital velocity is shown over-plotted with the best fit model and that of an aligned orbit (dotted line). The residuals are plotted below. }
\label{WASP24figure}
\end{center}
\end{figure*}

The measured parameters and uncertainties are given in Table \ref{Resultsw1}. Figure \ref{WASP1figure} shows the data over-plotted with the best-fit model, $\lambda$ = \lama, and the model corresponding to an aligned orbit, $\lambda$ = 0\degr. It is obvious that the amplitude and symmetry of the observations do not match with the aligned model. We conclude that WASP-1 has a severely misaligned orbit with respect to the rotation axis of the host star.  

The value of \vsini\ is constrained by the penalty function and thus the uncertainty in $\lambda$ is small despite the low impact parameter. It must be noted, however, that deriving \vsini\ from spectroscopic line broadening is susceptible to systematic errors due to the uncertainties in broadening mechanisms, such as micro- and macroturbulence, which are difficult to quantify. By altering \vsini\ by $\pm$ 1 \kms\ we find that the best-fit value of $\lambda$ changes by no more than 5\degr, so the interpretation of a highly obliquity is retained. The major factor leading the high precision in the determination of $\lambda$ is the system geometry. In a system with a low impact parameter and large misalignment, the form of the RM effect changes rapidly with $\lambda$ so it easy to differentiate between different angles. \citet{GW07} provide an relationship to estimate the expected uncertainty in $\lambda$. Substituting $\sigma_{\rm RV}$ = 8.5 m s$^{-1}$, $N_{\rm obs}$ = 12 and the derived values of $R_{\rm p}/R_{*}$, $b$, \vsini\ and $\lambda$ into their Equation 16, we obtain $\sigma_{\lambda}$ = 5.8\degr. This is in good agreement with the value we obtain and the small difference may due to the assumption that \vsini\ is a free parameter whereas here we constrain it with the penalty function.

\subsection{WASP-24}

A transit of WASP-24b was observed with the HARPS spectrograph on the night of 2010 April 10. We acquired 51 spectra covering the full transit (161 minutes, 22 observations) and a period of duration 212 minutes (28 observations) distributed evenly before and after transit. The atmospheric conditions were stable and the seeing was 0.7\arcsec. We also obtained 2 observations on the nights prior to and post transit to constrain the $\gamma$ velocity of the transit dataset. 

To fit the orbital parameters we obtained 8 HARPS observations between 2010 March 26 and 2010 March 28 and used 10 FIES and 18 CORALIE out-of-transit observations taken during the discovery of the planet \citep{Street10}. The HARPS observations are given in Table \ref{RVsW24}. 

\citet{Street10} reported that the orbit is not eccentric so $e$ was fixed to 0. The linear limb-darkening coefficient was chosen from the tables of \citet{Claret04} (ATLAS models) for the $V$ filter and fixed at $u = 0.66$. Values of the photometric parameters $P$, $T_{\rm 0}$,  $R_{\rm p}/R_{*}$, $a/R_{*}$ and $i_{\rm p}$ are taken from \citet{Street10} and are constrained through the $\chi^{2}$ fitting statistic below: 

\begin{eqnarray}
& \chi^2 & =  \sum_{i}^{} \left [ \frac{v_{i, \mathrm{obs}} - v_{i, \mathrm{calc}}}{\sigma_{i}} \right ]^{2}  + \;\;\;\nonumber \\
&  &     \left(  \frac{P - \periodvb ~{\rm d} + \periodeb~{\rm d} \times G(0,1)  }{\periodeb ~{\rm d}} \right)  ^{2}   + \;\;\;\nonumber \\
&  &    \left(  \frac{T_{\rm0} - \tzerovb +  \tzeroeb \times G(0,1)}{\tzeroeb} \right)  ^{2} +   \;\;\;\nonumber \\
&  &   \left(  \frac{R_{\rm p}/R_{*} - \rprsvb +  \rprseb  \times G(0,1)}{\rprseb} \right)  ^{2} + \;\;\;\nonumber \\ 
&  &      \left(  \frac{a/R_{*} - \arsvb + \arseb  \times G(0,1) }{\arseb} \right)  ^{2} + \;\;\;\nonumber \\
&  &    \left(  \frac{i_{\rm p} -  \incvb \degr +  \inceb \degr  \times G(0,1)}{\inceb \degr} \right)  ^{2}
\end{eqnarray}

Figure \ref{WASP24figure} shows that WASP-24 has a very symmetrical RM effect, moving from red shift to blue shift. This implies that that the planet moves in a prograde, well-aligned orbit. A fit to the observations indicates an obliquity to be $\lambda$ = \lamb. All the fitted parameters are shown in Table \ref{Resultsw24}. The fitted value of \vsini\ = \vsinib\ matches very well with that found from spectral line fitting, \vsini\ = 7.0 $\pm$ 1.0 \kms\ \citep{Street10}. WASP-24b has a high impact parameter ($b$ = 0.65) so a penalty function on \vsini was not needed. We found that adding such a penalty function constraining \vsini\ to the value from \citet{Street10} had no significant effect on the derived parameters.

\subsection{WASP-38}

\begin{figure*}
\begin{center}
\includegraphics[angle=0, width=\columnwidth]{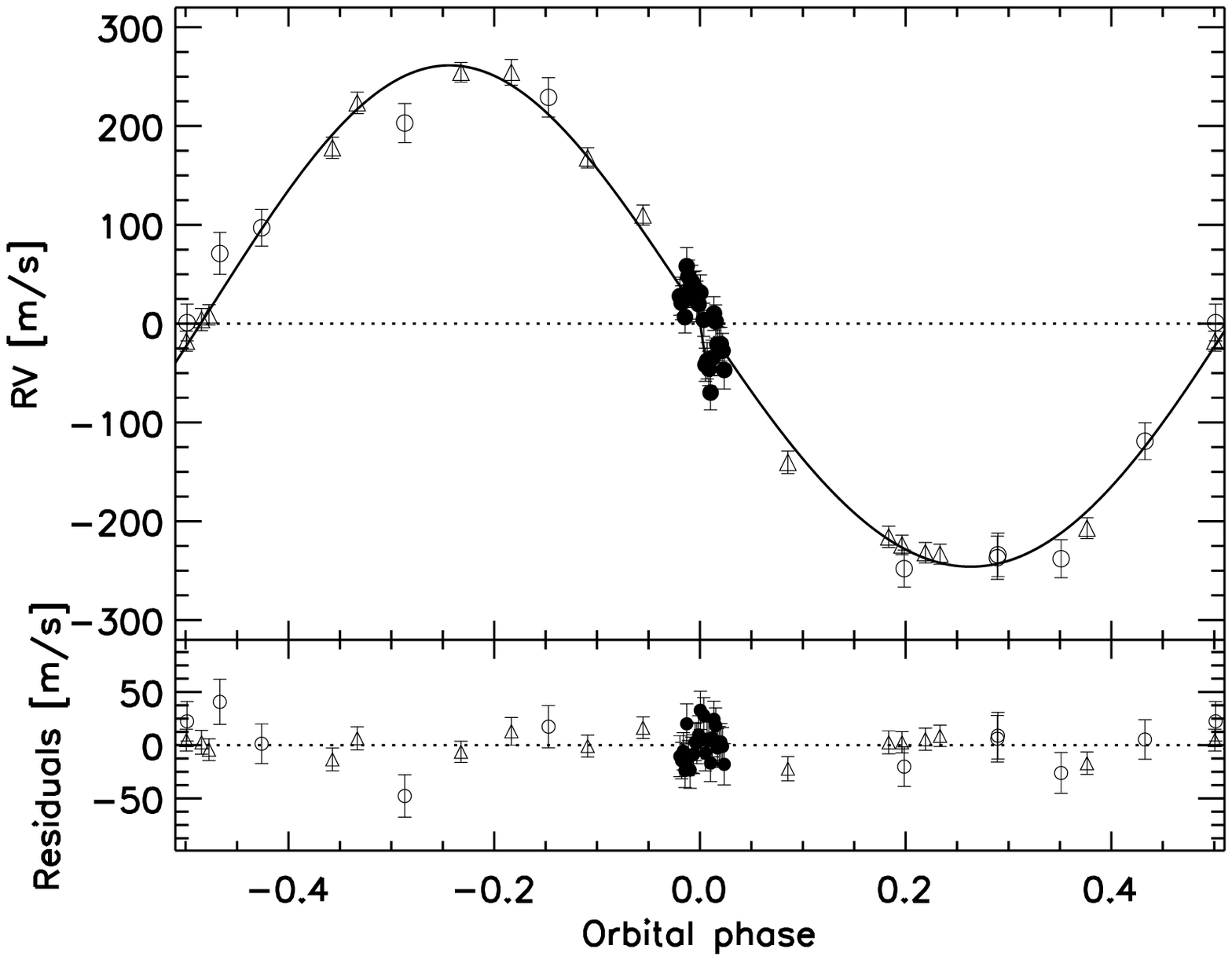}
\includegraphics[angle=0, width=\columnwidth]{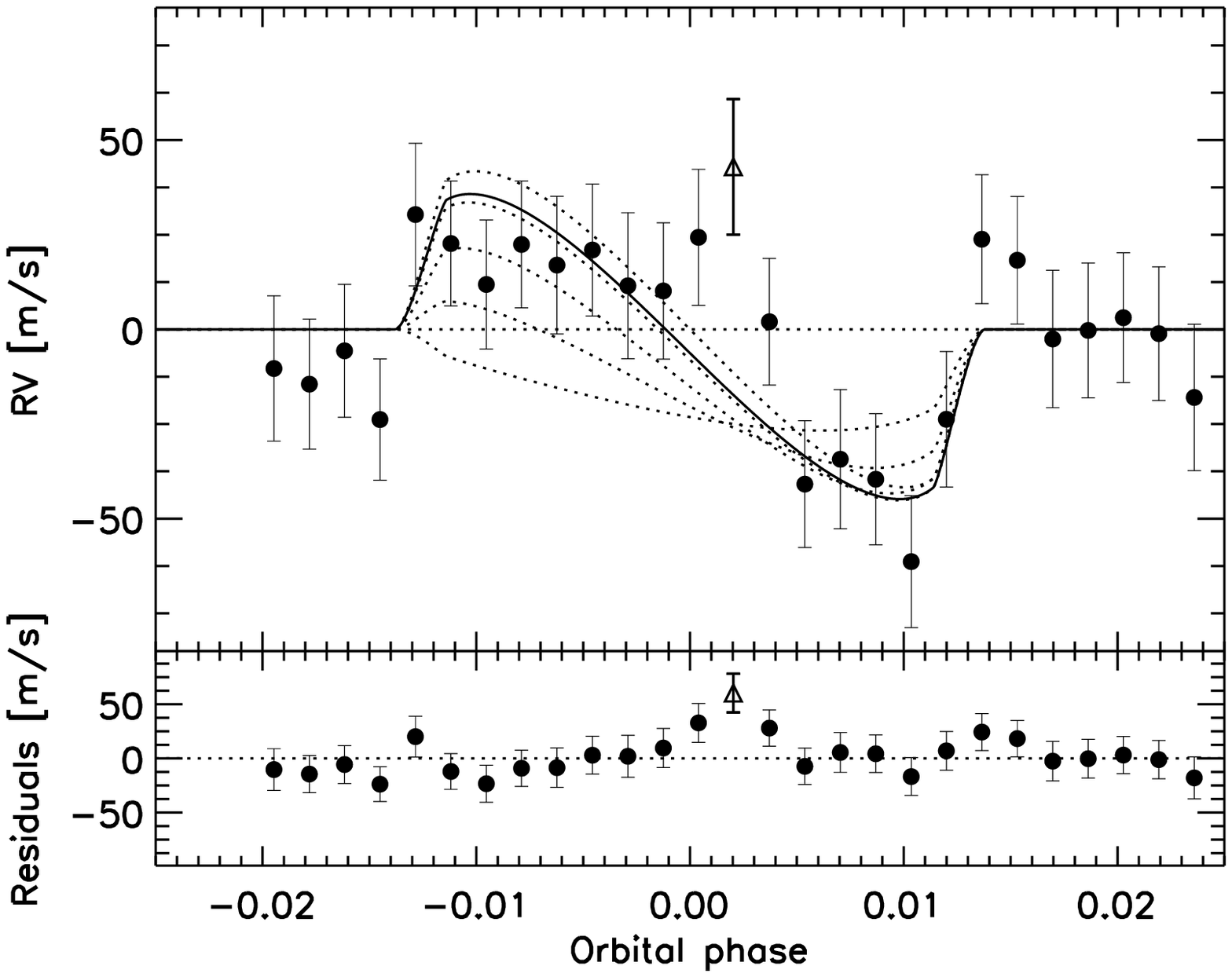}
\vspace{0.3cm}
\caption{Left: Phase folded radial velocities of WASP-38 over-plotted with the best-fit model. The systematic velocity has been removed (see Table \ref{Resultsw38}) and the residuals are shown below. The orbital points were taken with SOPHIE (open circles) and CORALE (open triangles) and the FIES transit sequence is shown in filled circles. Right: The spectroscopic transit is shown minus the orbital velocity and over-plotted with the best fit model. The point represented by an open triangle was omitted from the analysis (see text).  The dotted lines represent the models for $\lambda$ = 0\degr, 20\degr, 40\degr, 60\degr and 80\degr (from top to bottom) showing the change in the shape of the RM effect with $\lambda$. The residuals from the best-fit model are shown below. }
\label{WASP38figure}
\end{center}
\end{figure*}

The FIES spectrograph was employed to observe a transit of WASP-38 on 2010 June 8. We obtained 27 spectra during the night using an exposure time of 900s, giving 16 in-transit and 12 out-of-transit observations. The atmospheric conditions were stable and the seeing was 0.6\arcsec. To fit the orbit we used the SOPHIE and CORALIE out-of-transit data from \citet{Barros10}. The CORALIE radial velocity point at BJD = 2455404.6205 was removed, as it occurred during transit. To obtain $\chi^{2}_{red} = 1$, we required 15 and 8 m s$^{-1}$ to be added in quadrature to the SOPHIE and CORALIE internal uncertainties, respectively. 

\citet{Barros10} found that WASP-38 has a small but significantly eccentric orbit, so we fixed $e$ = 0.0321 and $\omega = -$19\degr, since those parameters have little effect on the fitted parameters once the transit ephemeris is specified \citep{Winn10hatp14}. The photometric parameters found from light curve fitting \citep{Barros10} are constrained through the penalty function shown in Equation \ref{eq38}. We determined the linear limb-darkening coefficient from the tables of \citet{Claret04} (ATLAS models) for the $V$ filter and fixed $u = 0.64$. 

The impact parameter of WASP-38 is low and has a relatively large uncertainty, $b$ = 0.27$^{+0.10}_{-0.14}$. As with WASP-1, we require an independent constraint on \vsini\ to break the degeneracy between \vsini\ and $\lambda$. The values of the photometric parameters and \vsini\ = 8.6 $\pm$ 0.4 \kms\ are taken from \citet{Barros10} and used in the $\chi^{2}$ statistic:

\begin{eqnarray}\label{eq38}
& \chi^2 &  =  \sum_{i}^{} \left [ \frac{v_{i, \mathrm{obs}} - v_{i, \mathrm{calc}}}{\sigma_{i}} \right ]^{2}  + \;\;\;\nonumber \\
&  &    \left(  \frac{v\sin{i} - \vsinivc\ \kms + \vsiniec ~ \kms  \times G(0,1) }{\vsiniec~ \kms} \right)  ^{2} + \;\;\;\nonumber \\
&  &     \left(  \frac{P - \periodvc ~{\rm d}  + \periodec~{\rm d}  \times G(0,1) }{\periodec ~{\rm d}} \right)  ^{2}   + \;\;\;\nonumber \\
&  &    \left(  \frac{T_{\rm0} - \tzerovc + \tzeroec  \times G(0,1)}{\tzeroec} \right)  ^{2} +   \;\;\;\nonumber \\
&  &   \left(  \frac{R_{\rm p}/R_{*} - \rprsvc + \rprsec  \times G(0,1)}{\rprsec} \right)  ^{2} + \;\;\;\nonumber \\ 
&  &      \left(  \frac{a/R_{*} - \arsvc +  \arsec  \times G(0,1) }{\arsec} \right)  ^{2} + \;\;\;\nonumber \\
&  &    \left(  \frac{i_{\rm p} -  \incvc \degr + \incec \degr  \times G(0,1)}{\incec \degr} \right)  ^{2}
\end{eqnarray}

Although the precision of the radial velocity measurements is lower than from the other spectrographs, we can see from the symmetry and characteristic red- then blue-shift of the RM effect that the planet is generally aligned and not retrograde. A fit to the data yields $\lambda$ = \lamc\ which is consistent with zero; the other fitted parameters are listed in Table \ref{Resultsw38}. The large uncertainty in $\lambda$ is due in part to the small differences in the shape of the RM effect when $\lambda$ and $b$ are small. Angles greater than approximately 60\degr\ produce a much more noticeable change in shape per degree than at low angles, as shown in Figure \ref{WASP38figure}. Thus the data allow us to rule out very high misalignment angles, however further observations are needed to reveal whether a small misalignment exists.

One observation lies more than three sigma from the best-fit model, see the open triangle in Figure \ref{WASP38figure}. We investigated whether this could be due to an incorrect wavelength calibration, but this does not appear to be the cause. However, seeing changes or guiding issues may be factors. The fit is improved by removing the point, so it was excluded in the analysis. The best-fit value of $\lambda$ obtained when the discrepant point is retained is $-42^{+32\tiny{\degr}}_{-13\tiny{\degr}}$ which is not inconsistent with zero at the 1.3$\sigma$ level.

\subsection{HAT-P-8}

We observed a transit of HAT-P-8 with the FIES spectrograph on  2010 August 31. An exposure time of 900s was used to obtain 14 in-transit and 14 out-of-transit spectra. The seeing was 0.55\arcsec\ and weather conditions were good. Between the sixth and seventh in-transit observations, the telescope was repointed when the rotator reached the maximum limit, which required $\sim$9 minutes but did not affect the quality of the dataset. We used the HIRES out-of-transit radial velocity points from \citet{Latham09} to constrain the orbit and fixed the eccentricity of the orbit to zero. The linear limb-darkening coefficient was set to the value found from the tables of \citet{Claret04} (ATLAS models) for the $V$ filter, $u = 0.69$.

\begin{figure}
\begin{center}
\includegraphics[angle=0, width=\columnwidth]{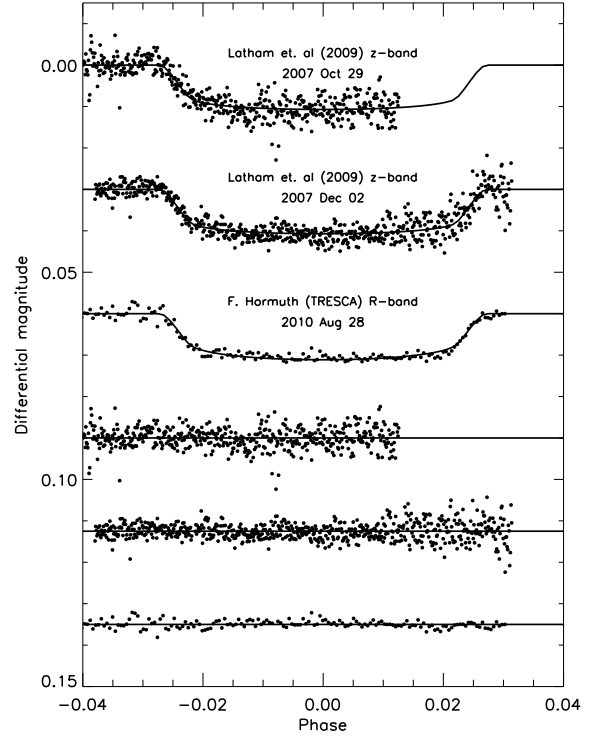}
\caption{Photometry of three transits of HAT-P-8 and the residuals from the best-fit model. }
\label{HATP8phot}
\end{center}
\end{figure}

Discovery light curves of HAT-P-8 were taken in 2007, so in order to improve the accuracy of the ephemeris, we fitted the two full z-band light curve presented in \citet{Latham09} and a high quality transit taken on 2010 August 28 by F. Harmuth (TRESCA/ETD), as shown in Figure \ref{HATP8phot}. The TRESCA observations were taken using the 1.2-m telescope at the Calar Alto Observatory in the R-band. As with WASP-1, an MCMC routine was used to fit the data and the parameters found for $P$, $T_{\rm 0}$,  $R_{\rm p}/R_{*}$, $a/R_{*}$ and $i_{\rm p}$ are shown in Equation \ref{eqw38}. 

An initial fit to the RM data found \vsini\ $\sim$ 16 \kms\ (and $\lambda$ $\sim$ 8\degr) to be significantly larger than the value of \vsini\ = 11.5 $\pm$ 0.5 \kms\ reported by \citet{Latham09}. Other similarly fast rotating stars have also been found to show a discrepancy between the value of \vsini\ derived from spectroscopic line broadening and from RM measurements \citep[e.g.][]{Winn07, Triaud09}. It is likely to be the result of the assumption that the line profile asymmetry caused by the planet blocking the rotating stellar surface can be modelled as a shift in the mean line position. For faster rotating stars, the asymmetry in the line profile is better resolved because of the broader line width, which causes a larger apparent shift than expected, as seen here. \citet{Hirano09} have addressed this issue by modifying the equations presented in \citet*{Ohta05} to compensate for this effect.  We implemented this solution as in \citet{Simpson10w3} and \citet{Bayliss10} by calculating the RM radial velocity shift as:  

\begin{equation}
 v_{\rm RM} = - \delta ~ v_{\rm p}  \left [1 + \frac{\sigma^{2}}{2\beta^{2}+\sigma^{2}}\right]^{3/2} \left [1 - \frac{v_{\rm p}^{2}}{2\beta^{2}+\sigma^{2}} \right]
\end{equation}

\noindent where $\beta$ is the intrinsic line width and $\sigma=\vsini/\alpha$, where $\alpha$ is a scaling factor depending on limb darkening parameters (see Equation F6 of H09). We determined $\beta$ as the quadrature sum of the macrotubulence ($v_{\rm mac}$), microturbulence ($v_{\rm mic}$) and instrumental profile. Following \citet{VF05} we used $v_{\rm mac}$ = 4.60 \kms\ and $v_{\rm mic}$ = 0.85 \kms. The instrumental profile is given by $c/R$ = 6.38 \kms, where $R$ is the instrumental resolution 46,000. Thus we determined $\beta$ = 7.9 \kms. We calculated $\alpha=1.31$ for the limb darkening coefficients $u_{1}=0.69$ and $u_{2}=0$. 

As with WASP-1b and WASP-38b, HAT-P-8b has a low impact parameter and therefore we placed a penalty function on the value of  \vsini\ found by \citet{Latham09} to constrain the fit.  The $\chi^{2}$ statistic used was:

\begin{eqnarray}\label{eqw38}
& \chi^2 & = \sum_{i}^{} \left [ \frac{v_{i, \mathrm{obs}} - v_{i, \mathrm{calc}}}{\sigma_{i}} \right ]^{2}  + \;\;\;\nonumber \\
&  &    \left(  \frac{v\sin{i} - \vsinivd\ \kms +  \vsinied ~ \kms  \times G(0,1) }{\vsinied~ \kms} \right)  ^{2} + \;\;\;\nonumber  \\
&  &     \left(  \frac{P - \periodvd ~{\rm d}  +\perioded~{\rm d}  \times G(0,1)}{\perioded ~{\rm d}} \right)  ^{2}   + \;\;\;\nonumber \\
&  &    \left(  \frac{T_{\rm0} - \tzerovd +  \tzeroed  \times G(0,1)}{\tzeroed} \right)  ^{2} +   \;\;\;\nonumber \\
&  &   \left(  \frac{R_{\rm p}/R_{*} - \rprsvd +  \rprsed  \times G(0,1)}{\rprsed} \right)  ^{2} + \;\;\;\nonumber \\ 
&  &      \left(  \frac{a/R_{*} - \arsvd +  \arsed  \times G(0,1) }{\arsed} \right)  ^{2} + \;\;\;\nonumber \\
&  &    \left(  \frac{i_{\rm p} -  \incvd \degr +  \inced \degr  \times G(0,1)}{\inced \degr} \right)  ^{2}
\end{eqnarray}

\begin{figure*}
\begin{center}
\includegraphics[angle=0, width=\columnwidth]{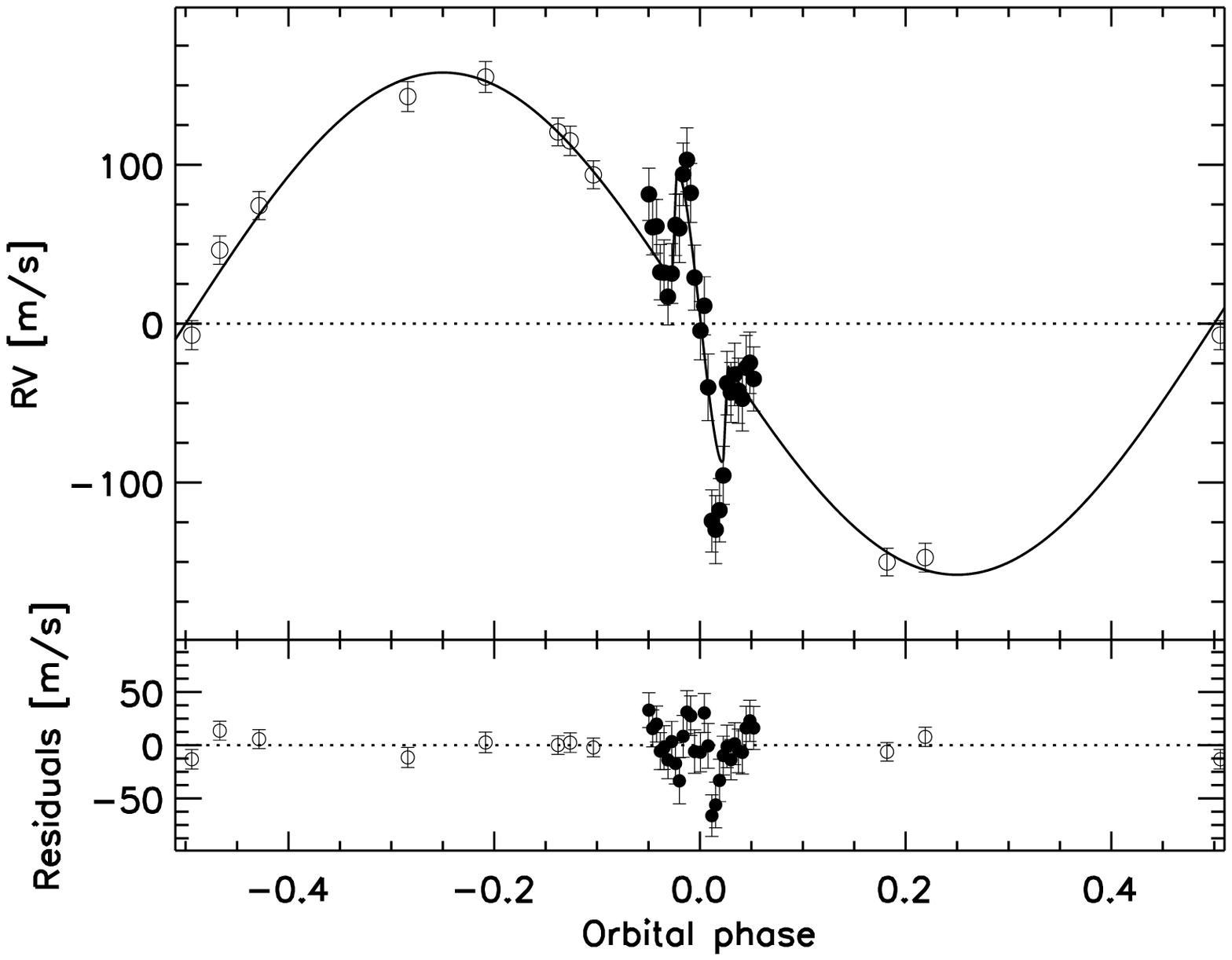}
\includegraphics[angle=0, width=\columnwidth]{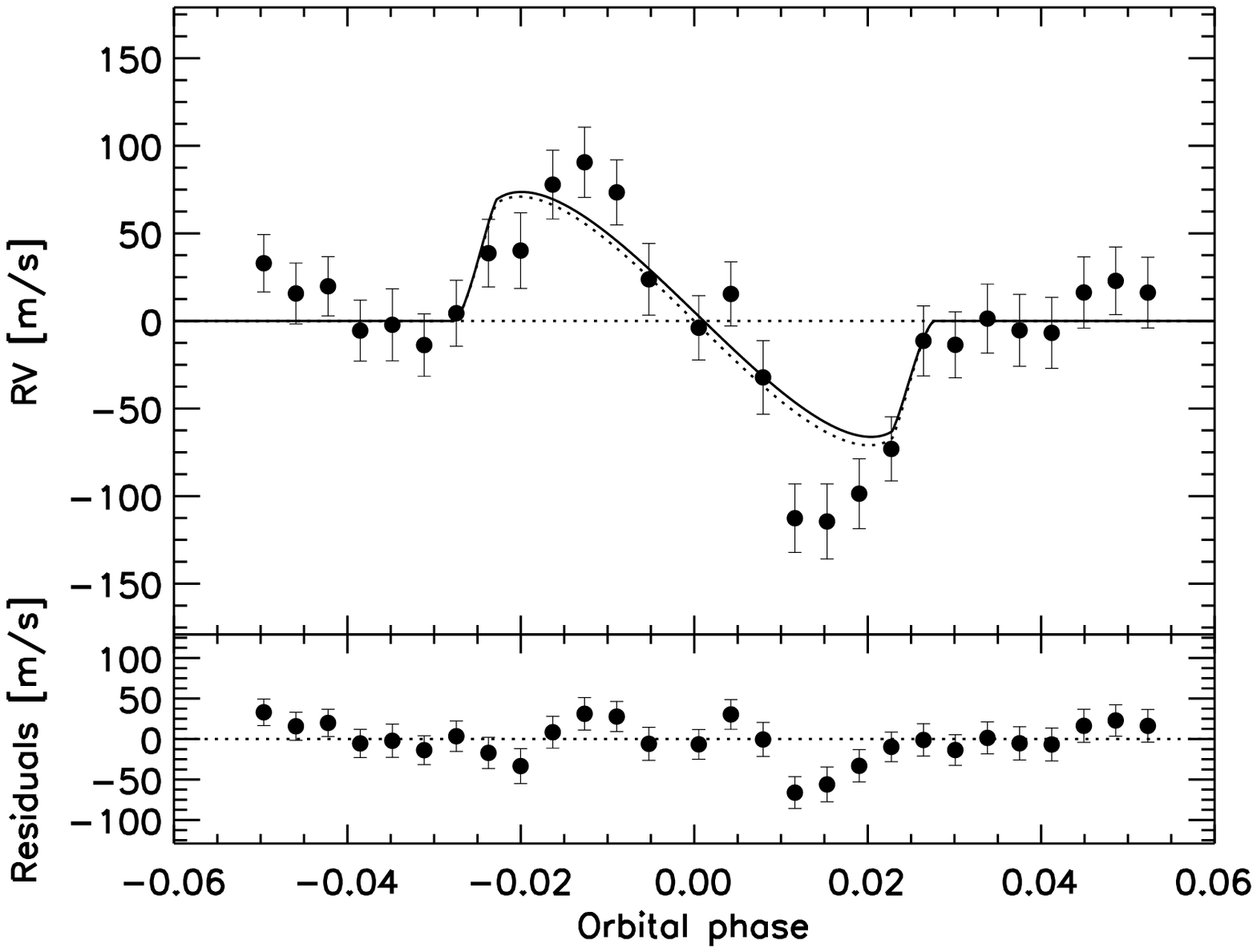}
\vspace{0.3cm}
\caption{Left: Phase-folded radial velocities of HAT-P-8 minus the systematic velocity (given in Table \ref{Resultshatp8}) and over-plotted with the best-fit model with residuals shown below. The orbital observations were taken using HIRES (open circles) and the transit sequence using FIES (filled circles). Bottom: The spectroscopic transit, minus the orbital velocity, is shown over-plotted with the best-fit model with the residuals shown below. A dotted line represents the RM effect of an aligned orbit. }
\label{HATP8figure}
\end{center}
\end{figure*}

From the form of the RM effect, HAT-P-8 appears to have a prograde aligned orbit, as shown in Figure \ref{HATP8figure}. A fit to the data yields $\lambda$ = \lamd, and the other fitted parameters are shown in Table \ref{Resultshatp8}. As with WASP-38b, the precision of FIES and the combination of low $b$ and $\lambda$ limits the accuracy of the result, but allows us to rule out highly misaligned orbits. A transit of HAT-P-8b was independently observed with SOPHIE (Moutou et al., in prep.); it provides results similar to the ones presented here.

Two observations show a large deviation from the best-fit model and again we could not determine an instrumental cause. We found that the fit was not significantly altered nor improved by removing the points so chose to retain them. The best-fit value of $\lambda$ having removed the two points (-18 $\pm$ 11\degr) is consistent with that found if the points are retained.

\section{Conclusions}\label{Discussion}

The spectroscopic transits of WASP-1b, WASP-24b, WASP-38b and HAT-P-8b have been observed using the HARPS, SOPHIE and FIES spectrographs. We modelled the RM effects and found the sky projected spin-orbit alignment angle of the systems to be $\lambda$ = \lama,  $\lambda$ = \lamb, $\lambda$ = \lamc\ and $\lambda$ = \lamd\ respectively. WASP-24b, WASP-38b and HAT-P-8b do not appear to be strongly misaligned and are consistent with zero within $2 \sigma$. This suggests they could have undergone a relatively non-violent migration process which did not perturb them from the primordial alignment of the proto-planetary disc. Alternatively, tidal interactions may have forced the stellar spin to align with the planetary orbit \citep{Winn10}. By contrast, WASP-1b joins the approximately one-third of planets with misaligned orbits and points towards a dynamically violent evolution such as gravitational scattering by other planets or the three-body Kozai mechanism. 

The uncertainties in $\lambda$ for HAT-P-8b and particularly WASP-38b are relatively large and allow for the possibility that the systems have a small misalignment. These larger uncertainties are due, in part, to the geometry of the systems. Both planets have low impact parameters and in this regime the shape of the RM effect, which largely determines $\lambda$, changes very subtly with $\lambda$ until angles of $\gtrsim$60\degr\ are reached. Thus, many values of $\lambda$ can be fit by the data and the uncertainty is large. This explains how we are able to obtain a much more precise value for WASP-1b which has a value of $| \lambda |$ which is much greater than zero. In addition, we are also limited by the instrumental noise on the FIES spectrograph and we recommend that further observations of these systems be undertaken to refine the parameters. 

We modelled the RM effect using the equations of \citet*{Ohta05}. However, it has been shown that this formulation does not estimate \vsini\ well for more rapidly rotating stars. We found this to be the case for HAT-P-8, which was measured to have \vsini\ = 11.5 $\pm$ 0.5  \kms\ from spectroscopic line broadening but a best-fit value of 16 \kms\ from the RM effect. For this case, we used the modified formulation of \citet{Hirano09} to model the data, and placed a constraint on \vsini\ using the value from spectroscopic line broadening to break the degeneracy between $\lambda$ and \vsini\ at low impact parameters. These modifications did not affect the conclusion that HAT-P-8b appears to be well aligned.   

The RM effect allows us to measure the spin-orbit angle in the plane of the sky. However, the line-of-sight alignment can be estimated by comparing the inclinations of the planetary orbit and stellar rotation axes $i_{\rm p}$ and $i_{*}$. Transiting planets naturally provide a tight constraint on $i_{p}$ given that the orbit must be close to edge-on in order for a eclipse to be observed. The stellar spin axis is more difficult to obtain and may be inferred by comparing the line-of-sight rotational velocity, \vsini, to the true stellar rotation velocity $v$. Theoretical predictions of $v$ are provided by \citet{Sman10} for many of the host stars of transiting planets. However, it can also be determined through the relation $v = 2 \pi R_{*} / P_{\rm rot}$, where $P_{\rm rot}$, the stellar rotation period, can be measured directly from spot modulation \citep[e.g.][]{Baliunas97, Simpson10rot} or indirectly from Ca II H \& K measurements \citep{Noyes84, Watson10sini} if this information is available.  

The four stars in this study have similar values of \teff\ (see Tables \ref{Resultsw1} -- \ref{Resultshatp8}) and age \citep[see][]{Sman10}, so we would expect them to have comparable values of $v$. \citet{Sman10} predicts this to be of the order of 7--9 \kms. In practice, we find a wide range of values of \vsini; \vsinia, \vsinib, \vsinic\ and \vsinid. This can be explained by supposing that the slower rotating stars are tilted so that we only measure a fraction of the true rotational velocity. WASP-1 is slower rotating than predicted (\vsini\ = 5.8 \kms, $v_{\rm sim}$ = 8.6 \kms) and from this, \citet{Sman10} deduced that the star and planet are significantly misaligned in the line-of-sight. This method yields four solutions for the inclination of the rotation axis of the star, $i_{*}$ $\sim$ 40\degr, 140\degr, 220\degr\ and 320\degr. 

The total alignment angle can be calculated through the following relation:

\begin{equation}
\cos{\psi} = \cos{i_{*}} \cos{i_{p}} + \sin{i_{*}} \sin{i_{p}} \cos{\lambda}.
\end{equation}

Substituting $i_{\rm p}$ = 88.6\degr\ and $\lambda$ = -79\degr, we obtain $\psi$ =  82\degr, 84\degr, 96\degr\ and 98\degr. We therefore conclude that WASP-1b has an almost polar orbit. \citet{Sman10} did not find WASP-24b and HAT-P-8b to have highly misaligned orbits in the line-of-sight, and WASP-38b has a similar \vsini, so we do not expect it to be misaligned either. This agrees with the lack of evidence for strong misalignments in the plane of the sky from the RM effect in these systems.  

Previous authors have noted correlations between spin-orbit misalignment and eccentric orbits and massive planets. With $e$ = 0.032 and $m_{\rm p}$ = 2.7 M$_{\rm J}$, WASP-38b may be expected to be misaligned, however we find no evidence for high spin-orbit angles. 

\citet{Winn10} noted that planets orbiting hot stars, defined as \teff $>$ 6250 K, tend to have misaligned orbits, whereas stars cooler than this host aligned planets. Low mass and long period planets are exceptions, as the timescale for tidal alignment is longer and may not have taken effect. All the planets in this study have temperatures placing them in the cool star category  and so WASP-24b, WASP-38b and HAT-P-8b appear to conform with this trend. However, with its $\sim$ 7 d orbit, WASP-38b has the potential to be an exception and to have retained its primordial spin-orbit angle.

By contrast, the WASP-1 system, with  \teff\ = 6110 $\pm$ 45 K, is unexpectedly misaligned. In this way, WASP-1b is similar to CoRoT-1b, which also orbits a cool star, \teff = 5950 $\pm$ 150 K \citep{Barge08}, and has a highly misaligned, almost polar orbit, $\lambda$ = 77 $\pm$ 11\degr\ \citep{Pont10} yet is not eccentric. Neither planets are low mass nor long period so do not meet the criteria to be exceptions from aligned orbits around cool stars. To bring the stars into the hot-star regime, the temperatures of CoRoT-1 and WASP-1 would have to be several $\sigma$ hotter than their best fit value, respectively. Whilst this could be the case, another possibility is that the distinction between cool and hot stars is not discrete and there may be a transition region between the two. 

Interestingly, \citet{Cameron07} reported a nearby stellar companion to WASP-1, and if associated, would have a projected orbital separation of $\sim$1800 AU. In this regime the Kozai mechanism can take effect and it is thought to significantly perturb planetary orbits. Although beyond the scope of this paper, it would be very informative to investigate whether the migration of WASP-1b could have been caused by interactions with a binary companion, if it is indeed associated.

The rapid increase in the number of measured spin-orbit angles, and the discovery of a large population of misaligned systems, has revolutionised our understanding of planetary migration. It is no longer believed that planet-disc interactions can fully explain the observed distribution of angles and other mechanisms must play a significant role. In this paper we have presented four new measurements of spin-orbit angles, and we find that WASP-1b has a near polar orbit, despite predictions that it would be aligned as it orbits a cool star. By contrast, WASP-38b is a massive planet on a moderately long period, eccentric orbit so has a high likelihood of misalignment but it does not show evidence for a large obliquity. In order to fully understand the processes which are at work, further measurements are needed to allow theories to be compared with a strong observational basis. \citet*{MJ10} estimate that a total of 40--100 measurements will be required to be confident in the mechanisms at work, and with the steady flow of new transiting planet discoveries, uncovering the mystery of migration may soon be within our reach.  

	\begin{table} 
	\centering
	\caption{Radial velocities of WASP-1 measured with SOPHIE during transit and FIES at other orbital phases.}
	\begin{tabular}{ccc}
	\hline \hline
	BJD		&RV		&Error	\\
	-2 400 000	&(\kms)	&(\kms)	\\
	\hline
	Planetary transit:\\
55099.33177 &     -13.2101 &        0.0083 \\
55099.34602 &     -13.2153 &        0.0086\\
55099.35950 &     -13.2128 &        0.0087\\
55099.37319 &     -13.2040 &        0.0086\\
55099.38674 &     -13.2146 &        0.0086\\
55099.40012 &     -13.2218 &        0.0085\\
55099.41520 &     -13.2341 &        0.0085\\
55099.42765 &     -13.2228 &        0.0085\\
55099.43928 &     -13.2454 &        0.0086\\
55099.45150 &     -13.2457 &        0.0087\\
55099.46349 &     -13.2614 &        0.0087\\
55099.47532 &     -13.2671 &        0.0088\\
55099.48773 &     -13.2527 &        0.0092\\
55099.50089 &     -13.2662 &        0.0088\\
55099.51269 &     -13.2596 &        0.0083\\
55099.52912 &     -13.2711 &        0.0072\\
55099.54277 &     -13.2774 &        0.0081\\
55099.55493 &     -13.2635 &        0.0086\\
55099.56709 &     -13.2843 &        0.0087\\
55099.58288 &     -13.2739 &        0.0088\\
55099.59579 &     -13.2883 &        0.0087\\
	\hline
		HJD		&RV		&Error	\\
	-2 400 000	&(\kms)	&(\kms)	\\
	\hline
	Other orbital phases:\\
54834.3378  &    -13.3780  &   0.0105 \\
54835.4419  &    -13.6246  &   0.0107\\
54836.4229  &    -13.4224  &  0.0096\\
55025.7036  &    -13.4250  &   0.0191\\
55041.6483  &    -13.5199  &   0.0249\\
55085.5576  &    -13.5266  &   0.0154\\
55086.5259  &    -13.4365  &   0.0393\\
55087.4126  &    -13.6088  &   0.0223\\
55096.5906  &    -13.4449  &   0.0215\\
 55097.6051 &     -13.6009 &    0.0181\\
55098.5769  &    -13.4039  &   0.0223\\
 55099.6120 &     -13.5478 &    0.0125\\
55100.5339  &    -13.5002  &   0.0102\\
  55119.4279  &-13.4457   &  0.0131\\
  55119.6984  &-13.5493   &  0.0187\\
	\hline
	\end{tabular}
	\label{RVsW1}
	\end{table}

	\begin{table}
	\centering
	\caption{Radial velocities of WASP-24 measured with HARPS during and outside transit.}
	\begin{tabular}{ccc}
	\hline \hline
	BJD		&RV		&Error	\\
	-2 400 000	&(\kms)	&(\kms)	\\
	\hline
	Planetary transit:\\
55296.6514 &     -17.7358  &       0.0059\\ 
55296.6651 &     -17.7463   &      0.0126 \\
55296.6701 &     -17.7492   &      0.0118 \\
55296.6750 &     -17.7407  &       0.0113 \\
55296.6801 &     -17.7504  &       0.0112 \\
55296.6850  &    -17.7733  &       0.0119 \\
55296.6900 &     -17.7487  &       0.0119 \\
55296.6950 &     -17.7436  &       0.0119 \\
55296.7000 &     -17.7492  &       0.0124 \\
55296.7051 &     -17.7629 &        0.0119 \\
55296.7100 &     -17.7551 &        0.0123 \\
55296.7150 &     -17.7619 &        0.0123 \\
55296.7201 &     -17.7668 &        0.0121 \\
55296.7250 &     -17.7521 &        0.0140 \\
55296.7300 &     -17.7364 &        0.0131 \\
55296.7350 &     -17.7379  &       0.0126 \\
55296.7400 &     -17.7434 &        0.0123 \\
55296.7450 &     -17.7512  &       0.0124 \\
55296.7501 &     -17.7645 &        0.0118 \\
55296.7550  &    -17.7403 &        0.0116 \\
55296.7600 &     -17.7633  &       0.0119 \\
55296.7650 &     -17.7531  &       0.0120 \\
55296.7700 &     -17.8006 &       0.0122 \\
55296.7751 &     -17.7885 &        0.0122 \\
55296.7800 &     -17.7813 &        0.0115 \\
55296.7851 &     -17.7965 &        0.0111 \\
55296.7901 &     -17.8128 &        0.0109 \\
55296.7950 &     -17.8160 &        0.0107 \\
55296.8000 &     -17.8399 &       0.0098 \\
55296.8049 &     -17.8265  &       0.0116 \\
55296.8100 &     -17.8495  &       0.0121 \\
55296.8150 &     -17.8366 &        0.0121 \\
55296.8199  &    -17.8141 &        0.0139 \\
55296.8251 &     -17.7952 &        0.0130 \\
55296.8300 &     -17.8240 &        0.0129 \\
55296.8350 &     -17.7943  &       0.0122 \\
55296.8401 &     -17.8418 &        0.0118 \\
55296.8451 &     -17.8177 &        0.0112 \\
55296.8500 &     -17.8012 &        0.0109 \\
55296.8550 &     -17.8154 &        0.0108 \\
55296.8600 &     -17.8018 &       0.0110 \\
55296.8650  &    -17.8293 &        0.0110 \\
55296.8701  &    -17.8289 &        0.0109 \\
55296.8750  &    -17.8200 &        0.0110 \\
55296.8801 &     -17.8355 &        0.0105 \\
55296.8851 &     -17.8203 &        0.0108 \\
55296.8901 &     -17.8294 &        0.0107 \\
55296.8951   &   -17.8398 &        0.0100 \\
55296.9000  &    -17.8602 &        0.0111 \\
55296.9051  &    -17.8438 &        0.0116 \\
55296.9101 &     -17.8287  &       0.0116 \\
\hline
		Other orbital phases:\\
55281.6933  &    -17.7475  &       0.0052 \\
55281.9080 &     -17.6632  &       0.0045 \\
55283.6945 &     -17.8572  &       0.0056 \\
55283.9139 &     -17.7825  &       0.0045 \\
55297.7382 &     -17.8696  &       0.0042 \\
55297.9076 &     -17.8057  &       0.0043 \\
55282.6762 &     -17.7674  &       0.0125\\
55282.6835 &     -17.7461  &       0.0118 \\
55282.6911 &     -17.7431  &       0.0118 \\
55282.6988 &     -17.7443 &        0.0141\\
55295.6485 &     -17.7747 &        0.0056 \\
55295.9092 &     -17.6802  &       0.0039 \\
	\hline
		\label{RVsW24}
	\end{tabular}
	\end{table}

		\begin{table} 
	\centering
	\caption{Radial velocities of WASP-38 measured with FIES during transit. The starred point (*) was omitted from the analysis}
	\begin{tabular}{ccccc}
	\hline \hline
	HJD		&RV		&Error	\\
	-2 400 000	&(\kms)	&(\kms)	\\
	\hline
       55356.4021  &	  -9.8108 &    0.0120\\
       55356.4135  &	  -9.8175 &    0.0084\\
       55356.4248  &	  -9.8113 &    0.0091\\
       55356.4362  &	  -9.8320 &    0.0057\\
       55356.4476  &	  -9.7805 &	0.0114\\
       55356.4590  &	  -9.7907 &    0.0069\\
       55356.4704  &	  -9.8042 &    0.0081\\
       55356.4817  &	  -9.7962 &    0.0074\\
       55356.4931  &	  -9.8042 &	0.0103\\
       55356.5045  &	  -9.8028 &    0.0089\\
       55356.5159  &	  -9.8149 &	0.0121\\
       55356.5273  &	  -9.8188 &    0.0010\\
       55356.5386  &	  -9.8072 &    0.0099\\
       55356.5500* &	  -9.7912 &    0.0098\\
       55356.5614  &	  -9.8347 &    0.0074\\
       55356.5728  &	  -9.8802 &    0.0075\\
       55356.5842  &	  -9.8762 &	0.0106\\
       55356.5956  &	  -9.8841 &    0.0087\\
       55356.6070  &	  -9.9084 &    0.0089\\
       55356.6183  &	  -9.8734 &    0.0098\\
       55356.6297  &	  -9.8284 &    0.0081\\
       55356.6411  &	  -9.8365 &    0.0076\\
       55356.6525  &	  -9.8599 &	0.0102\\
       55356.6639  &	  -9.8602 &    0.0096\\
       55356.6752  &	  -9.8593 &    0.0083\\
       55356.6866  &	  -9.8661 &    0.0093\\
       55356.6980  &	  -9.8855 &	0.0122\\
	\hline
	\end{tabular}
	\\
	\label{RVsW38}
	\end{table}
	
		\begin{table} 
	\centering
	\caption{Radial velocities of HAT-P-8 measured with FIES during transit.}
	\begin{tabular}{ccc}
	\hline \hline
	HJD		&RV		&Error	\\
	-2 400 000	&(\kms)	&(\kms)	\\
	\hline
       55440.4089  &	 -22.2836&    0.0067\\
       55440.4203 &	 -22.3043&    0.0087\\
       55440.4317 &	 -22.3038&    0.0078\\
       55440.4431 &	 -22.3326&    0.0088\\
       55440.4545 &	 -22.3329&     0.0141\\
       55440.4658 &	 -22.3481&    0.0096\\
       55440.4772 &	 -22.3335&     0.0114\\
       55440.4886 &	 -22.3029&     0.0121\\
       55440.5000 &	 -22.3050&     0.0156\\
       55440.5114 &	 -22.2710&     0.0127\\
       55440.5227 &	 -22.2619&     0.0134\\
       55440.5341 &	 -22.2827&     0.0109\\
       55440.5455 &	 -22.3360&     0.0139\\
       55440.5632 &	 -22.3694&     0.0106\\
       55440.5746 &	 -22.3538 &    0.0105\\
       55440.5860 &	 -22.4051 &    0.0147\\
       55440.5973 &	 -22.4891&     0.0126\\
       55440.6087 &	 -22.4947&     0.0153\\
       55440.6201 &	 -22.4824&     0.0132\\
       55440.6315 &	 -22.4606&     0.0104\\
       55440.6429 &	 -22.4025&     0.0132\\
       55440.6542 &	 -22.4083&     0.0114\\
       55440.6656 &	 -22.3970&     0.0128\\
       55440.6770 &	 -22.4072&     0.0140\\
       55440.6884 &	 -22.4123&     0.0137\\
       55440.6998 &	 -22.3928&     0.0137\\
       55440.7111 &	 -22.3896&     0.0122\\
       55440.7225 &	 -22.3998 &    0.0135\\
	\hline
	\end{tabular}
		\label{RVsHATP8}
	\end{table}

	\begin{table*}
	\caption{Derived system parameters and uncertainties for WASP-1. The effective temperature is taken from \citet{Stempels07}.}	
	\label{Resultsw1}
	\begin{tabular}{lc r@{}l}
	\hline\hline
	Parameter (units)			&Symbol & \multicolumn{2}{c}{Value}\\
	\hline
	Free parameters: & & \\[4pt]
	Projected alignment angle (\degr)				&$\lambda$		&\lamtva\,			&\lamtea				\\[4pt]
	RV semi-amplitude 	(\kms)					& $K$ 			&\ktva\, 			&$\pm$ \ktea			\\[4pt]
	Systemic velocity of SOPHIE transit dataset (\kms)	& $\gamma_1$		&\gammattva\,		&$\pm$ \gammattea		\\[4pt]
	Systemic velocity of SOPHIE orbital dataset (\kms)	& $\gamma_2$ 	&\gammaotva\,		&$\pm$ \gammaotea	\\[4pt]
	Systemic velocity of FIES orbital dataset (\kms)		& $\gamma_3$ 	&\gammaootva\,	&$\pm$ \gammaootea	\\[4pt]
	\hline
	Parameters controlled by priors: & &\\[4pt]
	Projected stellar rotation velocity (\kms)	& $v\sin{i}$ 		&\vsinitva\,	&$\pm$ \vsinitea     		\\[4pt]
	Period (days) 						&$P$			&\periodtva\,	&$\pm$ \periodtea		\\[4pt]
	Transit epoch	(HJD - 2 400 000) 				&$T_\mathrm{0}$	&\tzerotva\,	&$\pm$ \tzerotea		 \\[4pt]
	Planet/Star radius ratio				& $R_{p}/R_{*}$	&\rprstva\,		&$\pm$ \rprstea		\\[4pt]
	Scaled semi-major axis				&$a/R_{*}$ 		&\arstva\,		&$\pm$ \arstea			\\[4pt]
	Orbital inclination (\degr)				& $i$ 			&\inctva\,		&\inctea				\\[4pt]
	\hline
	Fixed parameters: & & \\[4pt]
	Eccentricity 				&$e$				& $0\,$ 				&\,				\\[4pt]
	Limb darkening 			&$u$				&$0.73\,$				&\,						 \\
	\hline
	Effective temperature (K)						& \teff\, 				& 6110\, & $\pm$ 45		\\
	\hline
	\vspace{0.2cm}
	\end{tabular}
	\end{table*}
	
	\begin{table*}
	\caption{Derived system parameters and uncertainties for WASP-24. The effective temperature is taken from \citet{Street10}.}	
	\label{Resultsw24}
	\begin{tabular}{lc r@{}l}
	\hline\hline
	Parameter (units)			&Symbol & \multicolumn{2}{c}{Value}\\
	\hline
	Free parameters: & & \\[4pt]
	Projected alignment angle (\degr)				&$\lambda$		&\lamtvb\,			&$\pm$ \lamteb		\\[4pt]
	Projected stellar rotation velocity (\kms)			& $v\sin{i}$ 		&\vsinitvb\,		&$\pm$ \vsiniteb     		\\[4pt]
	RV semi-amplitude 	(\kms)					& $K$ 			&\ktvb\, 			&$\pm$ \kteb			\\[4pt]
	Systemic velocity of HARPS transit dataset (\kms)	& $\gamma_1$		&\gammattvb\,		&$\pm$ \gammatteb		\\[4pt]
	Systemic velocity of HARPS orbital dataset (\kms)	& $\gamma_2$ 	&\gammaoootvb\,	&$\pm$ \gammaoooteb	\\[4pt]
	Systemic velocity of FIES orbital dataset (\kms)		& $\gamma_3$ 	&\gammaootvb\,	&$\pm$ \gammaooteb	\\[4pt]
	Systemic velocity of CORALIE orbital dataset (\kms)	& $\gamma_4$ 	&\gammaotvb\,		&$\pm$ \gammaoteb	\\[4pt]
	\hline
	Parameters controlled by priors: & &\\[4pt]

	Period (days) 						&$P$			&\periodtvb\,	&$\pm$ \periodteb		\\[4pt]
	Transit epoch	(HJD - 2 400 000) 				&$T_\mathrm{0}$	&\tzerotvb\,	&$\pm$ \tzeroteb		 \\[4pt]
	Planet/Star radius ratio				& $R_{p}/R_{*}$	&\rprstvb\,		&$\pm$ \rprsteb		\\[4pt]
	Scaled semi-major axis				&$a/R_{*}$ 		&\arstvb\,		&$\pm$ \arsteb		\\[4pt]
	Orbital inclination (\degr)				& $i$ 			&\inctvb\,		&$\pm$ \incteb				\\[4pt]
	\hline
	Fixed parameters: & & \\[4pt]
	Eccentricity 				&$e$				& $0\,$ 				&\,				\\[4pt]
	Limb darkening 			&$u$				&$0.66\,$				&\,						 \\
	\hline
	Effective temperature (K)						& \teff\, 				& 6075\, & $\pm$ 100		\\
	\hline
	\end{tabular}
	\end{table*}

	\begin{table*}
	\caption{Derived system parameters and uncertainties for WASP-38. The effective temperature is taken from \citet{Barros10}.}	
	\label{Resultsw38}
	\begin{tabular}{lc r@{}l}
	\hline\hline
	Parameter (units)			&Symbol & \multicolumn{2}{c}{Value}\\
	\hline
	Projected alignment angle (\degr)				&$\lambda$		&\lamtvc\,			&\lamtec				\\[4pt]
	RV semi-amplitude 	(\kms)					& $K$ 			&\ktvc\, 			&$\pm$ \ktec			\\[4pt]
	Systemic velocity of FIES transit dataset (\kms)		& $\gamma_1$		&\gammattvc\,		&\gammattec			\\[4pt]
	Systemic velocity of SOPHIE orbital dataset (\kms)	& $\gamma_2$ 	&\gammaotvc\,		&$\pm$ \gammaotec		\\[4pt]
	Systemic velocity of CORALIE orbital dataset (\kms)	& $\gamma_3$ 	&\gammaootvc\,	&$\pm$ \gammaootec	\\[4pt]
	\hline
	Parameters controlled by priors: & &\\[4pt]
	Projected stellar rotation velocity (\kms)	& $v\sin{i}$ 		&\vsinitvc\,	&$\pm$ \vsinitec    		\\[4pt]
	Period (days) 						&$P$			&\periodtvc\,	&$\pm$ \periodtec		\\[4pt]
	Transit epoch	(HJD - 2 400 000) 				&$T_\mathrm{0}$	&\tzerotvc\,	&$\pm$ \tzerotec		 \\[4pt]
	Planet/Star radius ratio				& $R_{p}/R_{*}$	&\rprstvc\,		&$\pm$ \rprstec		\\[4pt]
	Scaled semi-major axis				&$a/R_{*}$ 		&\arstvc\,		&$\pm$ \arstec			\\[4pt]
	Orbital inclination (\degr)				& $i$ 			&\inctvc\,		&\inctea				\\[4pt]
	\hline
	Fixed parameters: & & \\[4pt]
	Eccentricity			&$e$				&$0.032$ 				&\,					\\[4pt]
	Longitude of periastron	&$\omega$			&$-19.0$ 				&\,					\\[4pt]
	Limb darkening		&$u$				&$0.64\,$				&\,					\\[4pt]
	\hline
	Effective temperature (K)						& \teff\, 				& 6150\, & $\pm$ 80		\\
	\hline
	\vspace{0.2cm}
	\end{tabular}
	\end{table*}
	
		\begin{table*}
	\caption{Derived system parameters and uncertainties for HAT-P-8. The effective temperature is taken from \citet{Latham09}}	
	\label{Resultshatp8}
	\begin{tabular}{lc r@{}l}
	\hline\hline
	Parameter (units)			&Symbol & \multicolumn{2}{c}{Value}\\
	\hline
	Free parameters: & & \\[4pt]
	Projected alignment angle (\degr)				&$\lambda$		&\lamtvd\,			&\lamted				\\[4pt]
	RV semi-amplitude 	(\kms)					& $K$ 			&\ktvd\, 			&$\pm$ \kted			\\[4pt]
	Systemic velocity of FIES transit dataset (\kms)		& $\gamma_1$		&\gammattvd\,		&$\pm$ \gammatted		\\[4pt]
	Systemic velocity of HIRES orbital dataset (\kms)	& $\gamma_2$ 	&\gammaotvd\,		&$\pm$ \gammaoted	\\[4pt]
	\hline
	Parameters controlled by priors: & &\\[4pt]
	Projected stellar rotation velocity (\kms)	& $v\sin{i}$ 		&\vsinitvd\,	&$\pm$ \vsinited     		\\[4pt]
	Period (days) 						&$P$			&\periodtvd\,	&$\pm$ \periodted		\\[4pt]
	Transit epoch	(HJD - 2 400 000) 				&$T_\mathrm{0}$	&\tzerotvd\,	&$\pm$ \tzeroted		 \\[4pt]
	Planet/Star radius ratio				& $R_{p}/R_{*}$	&\rprstvd\,		&$\pm$ \rprsted		\\[4pt]
	Scaled semi-major axis				&$a/R_{*}$ 		&\arstvd\,		&\arsted				\\[4pt]
	Orbital inclination (\degr)				& $i$ 			&\inctvd\,		&\incted				\\[4pt]
	\hline
	Fixed parameters: & & \\[4pt]
	Eccentricity			&$e$				&$0\,$ 				&\,					\\[4pt]
	Limb darkening		&$u$				&$0.69\,$				&\,						 \\[4pt]
	\hline
	Effective temperature (K)	& \teff\, 			& 6200\,	 & $\pm$ 80		\\
	\hline
	\end{tabular}
	\end{table*}

\section*{Acknowledgments}

We thank J. Southworth for making available his code, JKTLD, for calculating theoretical limb darkening coefficients. FPK is grateful to AWE Aldermaston for the award of a William Penny Fellowship.

\bibliographystyle{aa.bst}
\bibliography{bib.bib}

\begin{thebibliography}{65}
\expandafter\ifx\csname natexlab\endcsname\relax\def\natexlab#1{#1}\fi

\bibitem[{{Baliunas} {et~al.}(1997){Baliunas}, {Henry}, {Donahue}, {Fekel}, \&
  {Soon}}]{Baliunas97}
{Baliunas}, S.~L., {Henry}, G.~W., {Donahue}, R.~A., {Fekel}, F.~C., \& {Soon},
  W.~H. 1997, ApJL, 474, L119+

\bibitem[{{Baranne} {et~al.}(1996){Baranne}, {Queloz}, {Mayor}, {Adrianzyk},
  {Knispel}, \& {et al.}}]{Baranne96}
{Baranne}, A., {Queloz}, D., {Mayor}, M., {et~al.} 1996, A\&ASS, 119, 373

\bibitem[{{Barge} {et~al.}(2008){Barge}, {Baglin}, {Auvergne}, {Rauer},
  {L{\'e}ger}, {Schneider}, {Pont}, {Aigrain}, {Almenara}, {Alonso},
  {Barbieri}, {Bord{\'e}}, {Bouchy}, {Deeg}, {La Reza}, {Deleuil}, {Dvorak},
  {Erikson}, {Fridlund}, {Gillon}, {Gondoin}, {Guillot}, {Hatzes}, {Hebrard},
  {Jorda}, {Kabath}, {Lammer}, {Llebaria}, {Loeillet}, {Magain}, {Mazeh},
  {Moutou}, {Ollivier}, {P{\"a}tzold}, {Queloz}, {Rouan}, {Shporer}, \&
  {Wuchterl}}]{Barge08}
{Barge}, P., {Baglin}, A., {Auvergne}, M., {et~al.} 2008, A\&A, 482, L17

\bibitem[{{Barros} {et~al.}(2010){Barros}, {Faedi}, {Collier Cameron},
  {Lister}, {McCormac}, {Pollacco}, {Simpson}, {Smalley}, {Street}, {Todd},
  {Triaud}, {Boisse}, {Bouchy}, {Hebrard}, {Moutou}, {Pepe}, {Queloz},
  {Santerne}, {Segransan}, {Udry}, {Bento}, {Butters}, {Enoch}, {Haswell},
  {Hellier}, {Keenan}, {Miller}, {Moulds}, {Norton}, {Parley}, {Skillen},
  {Watson}, {West}, \& {Wheatley}}]{Barros10}
{Barros}, S.~C.~C., {Faedi}, F., {Collier Cameron}, A., {et~al.} 2010, ArXiv
  e-prints

\bibitem[{{Bate} {et~al.}(2010){Bate}, {Lodato}, \& {Pringle}}]{Bate10}
{Bate}, M.~R., {Lodato}, G., \& {Pringle}, J.~E. 2010, MNRAS, 401, 1505

\bibitem[{{Bayliss} {et~al.}(2010){Bayliss}, {Winn}, {Mardling}, \&
  {Sackett}}]{Bayliss10}
{Bayliss}, D.~D.~R., {Winn}, J.~N., {Mardling}, R.~A., \& {Sackett}, P.~D.
  2010, ApJL, 722, L224

\bibitem[{{Bouchy} {et~al.}(2009){Bouchy}, {H{\'e}brard}, {Udry}, {Delfosse},
  \& {Boisse}}]{Bouchy09}
{Bouchy}, F., {H{\'e}brard}, G., {Udry}, S., {Delfosse}, X., \& {Boisse}, I.~e.
  2009, A\&A, 505, 853

\bibitem[{{Buchhave} {et~al.}(2010){Buchhave}, {Bakos}, {Hartman}, {Torres},
  {Kov{\'a}cs}, {Latham}, {Noyes}, {Esquerdo}, {Everett}, {Howard}, {Marcy},
  {Fischer}, {Johnson}, {Andersen}, {F{\H u}r{\'e}sz}, {Perumpilly},
  {Sasselov}, {Stefanik}, {B{\'e}ky}, {L{\'a}z{\'a}r}, {Papp}, \&
  {S{\'a}ri}}]{Buchhave10}
{Buchhave}, L.~A., {Bakos}, G.~{\'A}., {Hartman}, J.~D., {et~al.} 2010, ApJ,
  720, 1118

\bibitem[{{Butler} \& {Marcy}(1996)}]{BM96}
{Butler}, R.~P. \& {Marcy}, G.~W. 1996, ApJL, 464, L153+

\bibitem[{{Cameron} {et~al.}(2007){Cameron}, {Bouchy}, {H{\'e}brard}, {Maxted},
  {Pollacco}, \& {et al.}}]{Cameron07}
{Cameron}, A.~C., {Bouchy}, F., {H{\'e}brard}, G., {et~al.} 2007, MNRAS, 375,
  951

\bibitem[{{Charbonneau} {et~al.}(2007){Charbonneau}, {Winn}, {Everett},
  {Latham}, {Holman}, {Esquerdo}, \& {O'Donovan}}]{Charbonneau07}
{Charbonneau}, D., {Winn}, J.~N., {Everett}, M.~E., {et~al.} 2007, ApJ, 658,
  1322

\bibitem[{{Claret}(2004)}]{Claret04}
{Claret}, A. 2004, A\&A, 428, 1001

\bibitem[{{Fabrycky} \& {Winn}(2009)}]{FW09}
{Fabrycky}, D.~C. \& {Winn}, J.~N. 2009, ApJ, 696, 1230

\bibitem[{{Gaudi} \& {Winn}(2007)}]{GW07}
{Gaudi}, B.~S. \& {Winn}, J.~N. 2007, ApJ, 655, 550

\bibitem[{{Gillon}(2009)}]{Gillon09}
{Gillon}, M. 2009, arXiv:0906.4904

\bibitem[{{Hebrard} {et~al.}(2008){Hebrard}, {Bouchy}, {Pont}, {Loeillet},
  {Rabus}, \& {et al.}}]{Hebrard08}
{Hebrard}, G., {Bouchy}, F., {Pont}, F., {et~al.} 2008, A\&A, 488, 763

\bibitem[{{H{\'e}brard} {et~al.}(2010){H{\'e}brard}, {D{\'e}sert},
  {D{\'{\i}}az}, {Boisse}, {Bouchy}, {Lecavelier Des Etangs}, {Moutou},
  {Ehrenreich}, {Arnold}, {Bonfils}, {Delfosse}, {Desort}, {Eggenberger},
  {Forveille}, {Gregorio}, {Lagrange}, {Lovis}, {Pepe}, {Perrier}, {Pont},
  {Queloz}, {Santerne}, {Santos}, {S{\'e}gransan}, {Sing}, {Udry}, \&
  {Vidal-Madjar}}]{Hebrard10}
{H{\'e}brard}, G., {D{\'e}sert}, J., {D{\'{\i}}az}, R.~F., {et~al.} 2010, A\&A,
  516, A95+

\bibitem[{{Hirano} {et~al.}(2009){Hirano}, {Suto}, {Taruya}, {Narita}, {Sato},
  \& {et al.}}]{Hirano09}
{Hirano}, T., {Suto}, Y., {Taruya}, A., {et~al.} 2009, arxiv:0910.2365

\bibitem[{{Johnson} {et~al.}(2009){Johnson}, {Winn}, {Albrecht}, {Howard},
  {Marcy}, \& {et al.}}]{Johnson09}
{Johnson}, J.~A., {Winn}, J.~N., {Albrecht}, S., {et~al.} 2009, PASP, 121, 1104

\bibitem[{{Lai} {et~al.}(2010){Lai}, {Foucart}, \& {Lin}}]{Lai10}
{Lai}, D., {Foucart}, F., \& {Lin}, D.~N.~C. 2010, ArXiv e-prints

\bibitem[{{Latham} {et~al.}(2009){Latham}, {Bakos}, {Torres}, {Stefanik},
  {Noyes}, {Kov{\'a}cs}, {P{\'a}l}, {Marcy}, {Fischer}, {Butler}, {Sip{\H
  o}cz}, {Sasselov}, {Esquerdo}, {Vogt}, {Hartman}, {Kov{\'a}cs},
  {L{\'a}z{\'a}r}, {Papp}, \& {S{\'a}ri}}]{Latham09}
{Latham}, D.~W., {Bakos}, G.~{\'A}., {Torres}, G., {et~al.} 2009, ApJ, 704,
  1107

\bibitem[{{Lin} {et~al.}(1996){Lin}, {Bodenheimer}, \& {Richardson}}]{Lin96}
{Lin}, D.~N.~C., {Bodenheimer}, P., \& {Richardson}, D.~C. 1996, Nature, 380,
  606

\bibitem[{{Loeillet} {et~al.}(2008){Loeillet}, {Shporer}, {Bouchy}, {Pont},
  {Mazeh}, {Beuzit}, {Boisse}, {Bonfils}, {da Silva}, {Delfosse}, {Desort},
  {Ecuvillon}, {Forveille}, {Galland}, {Gallenne}, {H{\'e}brard}, {Lagrange},
  {Lovis}, {Mayor}, {Moutou}, {Pepe}, {Perrier}, {Queloz}, {S{\'e}gransan},
  {Sivan}, {Santos}, {Tsodikovich}, {Udry}, \& {Vidal-Madjar}}]{Loeillet08}
{Loeillet}, B., {Shporer}, A., {Bouchy}, F., {et~al.} 2008, A\&A, 481, 529

\bibitem[{{Lovis} \& {Pepe}(2007)}]{LovisPepe07}
{Lovis}, C. \& {Pepe}, F. 2007, A\&A, 468, 1115

\bibitem[{{Marcy} \& {Butler}(1996)}]{MB96}
{Marcy}, G.~W. \& {Butler}, R.~P. 1996, ApJL, 464, L147+

\bibitem[{{Markwardt}(2009)}]{mpfit}
{Markwardt}, C.~B. 2009, in Astronomical Society of the Pacific Conference
  Series, Vol. 411, Astronomical Society of the Pacific Conference Series, ed.
  {D.~A.~Bohlender, D.~Durand, \& P.~Dowler}, 251

\bibitem[{{Mayor} {et~al.}(2003){Mayor}, {Pepe}, {Queloz}, {Bouchy},
  {Rupprecht}, {Lo Curto}, {Avila}, {Benz}, {Bertaux}, {Bonfils}, {Dall},
  {Dekker}, {Delabre}, {Eckert}, {Fleury}, {Gilliotte}, {Gojak}, {Guzman},
  {Kohler}, {Lizon}, {Longinotti}, {Lovis}, {Megevand}, {Pasquini}, {Reyes},
  {Sivan}, {Sosnowska}, {Soto}, {Udry}, {van Kesteren}, {Weber}, \&
  {Weilenmann}}]{Mayor03}
{Mayor}, M., {Pepe}, F., {Queloz}, D., {et~al.} 2003, The Messenger, 114, 20

\bibitem[{{Mayor} \& {Queloz}(1995)}]{MQ95}
{Mayor}, M. \& {Queloz}, D. 1995, Nature, 378, 355

\bibitem[{{McLaughlin}(1924)}]{McLaughlin24}
{McLaughlin}, D.~B. 1924, ApJ, 60, 22

\bibitem[{{Morton} \& {Johnson}(2010)}]{MJ10}
{Morton}, T.~D. \& {Johnson}, J.~A. 2010, ArXiv e-prints

\bibitem[{{Moutou} {et~al.}(2009){Moutou}, {H{\'e}brard}, {Bouchy},
  {Eggenberger}, \& {Boisse}}]{Moutou09}
{Moutou}, C., {H{\'e}brard}, G., {Bouchy}, F., {Eggenberger}, A., \& {Boisse},
  I. 2009, A\&A, 498, L5

\bibitem[{{Murray} {et~al.}(1998){Murray}, {Hansen}, {Holman}, \&
  {Tremaine}}]{Murray98}
{Murray}, N., {Hansen}, B., {Holman}, M., \& {Tremaine}, S. 1998, Science, 279,
  69

\bibitem[{{Nagasawa} {et~al.}(2008){Nagasawa}, {Ida}, \& {Bessho}}]{Nagasawa08}
{Nagasawa}, M., {Ida}, S., \& {Bessho}, T. 2008, ApJ, 678, 498

\bibitem[{{Narita} {et~al.}(2009){Narita}, {Hirano}, {Sato}, {Winn}, {Suto},
  {Turner}, {Aoki}, {Tamura}, \& {Yamada}}]{Narita09hd17156}
{Narita}, N., {Hirano}, T., {Sato}, B., {et~al.} 2009, PASJ, 61, 991

\bibitem[{{Noyes} {et~al.}(1984){Noyes}, {Hartmann}, {Baliunas}, {Duncan}, \&
  {Vaughan}}]{Noyes84}
{Noyes}, R.~W., {Hartmann}, L.~W., {Baliunas}, S.~L., {Duncan}, D.~K., \&
  {Vaughan}, A.~H. 1984, ApJ, 279, 763

\bibitem[{{Ohta} {et~al.}(2005){Ohta}, {Taruya}, \& {Suto}}]{Ohta05}
{Ohta}, Y., {Taruya}, A., \& {Suto}, Y. 2005, ApJ, 622, 1118

\bibitem[{{Pepe} {et~al.}(2002){Pepe}, {Mayor}, {Galland}, {Naef}, {Queloz}, \&
  {et al.}}]{Pepe02}
{Pepe}, F., {Mayor}, M., {Galland}, F., {et~al.} 2002, A\&A, 388, 632

\bibitem[{{Perruchot} {et~al.}(2008){Perruchot}, {Kohler}, {Bouchy}, {Richaud},
  \& {Richaud}}]{Perruchot08}
{Perruchot}, S., {Kohler}, D., {Bouchy}, F., {Richaud}, Y., \& {Richaud}, P.~e.
  2008, in Presented at the Society of Photo-Optical Instrumentation Engineers
  (SPIE) Conference, Vol. 7014, Society of Photo-Optical Instrumentation
  Engineers (SPIE) Conference Series

\bibitem[{{Poddan{\'y}} {et~al.}(2010){Poddan{\'y}}, {Br{\'a}t}, \&
  {Pejcha}}]{Poddany10}
{Poddan{\'y}}, S., {Br{\'a}t}, L., \& {Pejcha}, O. 2010, NewA, 15, 297

\bibitem[{{Pollacco} {et~al.}(2008){Pollacco}, {Skillen}, {Collier Cameron},
  {Loeillet}, {Stempels}, \& {et al.}}]{Pollacco08}
{Pollacco}, D., {Skillen}, I., {Collier Cameron}, A., {et~al.} 2008, MNRAS,
  385, 1576

\bibitem[{{Pont} {et~al.}(2010){Pont}, {Endl}, {Cochran}, {Barnes}, {Sneden},
  {MacQueen}, {Moutou}, {Aigrain}, {Alonso}, {Baglin}, {Bouchy}, {Deleuil},
  {Fridlund}, {H{\'e}brard}, {Hatzes}, {Mazeh}, \& {Shporer}}]{Pont10}
{Pont}, F., {Endl}, M., {Cochran}, W.~D., {et~al.} 2010, MNRAS, 402, L1

\bibitem[{{Pont} {et~al.}(2009){Pont}, {H{\'e}brard}, {Irwin}, {Bouchy},
  {Moutou}, \& {et~al.}}]{Pont09b}
{Pont}, F., {H{\'e}brard}, G., {Irwin}, J.~M., {et~al.} 2009, A\&A, 502, 695

\bibitem[{{Queloz} {et~al.}(2010){Queloz}, {Anderson}, {Collier Cameron},
  {Gillon}, {Hebb}, {Hellier}, {Maxted}, {Pepe}, {Pollacco}, {S{\'e}gransan},
  {Smalley}, {Triaud}, {Udry}, \& {West}}]{Queloz10}
{Queloz}, D., {Anderson}, D., {Collier Cameron}, A., {et~al.} 2010, A\&A, 517,
  L1+

\bibitem[{{Queloz} {et~al.}(1999){Queloz}, {Casse}, \& {Mayor}}]{Queloz99}
{Queloz}, D., {Casse}, M., \& {Mayor}, M. 1999, in Astronomical Society of the
  Pacific Conference Series, Vol. 185, IAU Colloq. 170: Precise Stellar Radial
  Velocities, ed. {J.~B.~Hearnshaw \& C.~D.~Scarfe}, 13--+

\bibitem[{{Rasio} \& {Ford}(1996)}]{RF96}
{Rasio}, F.~A. \& {Ford}, E.~B. 1996, Science, 274, 954

\bibitem[{{Rossiter}(1924)}]{Rossiter24}
{Rossiter}, R.~A. 1924, ApJ, 60, 15

\bibitem[{{Rupprecht} {et~al.}(2004){Rupprecht}, {Pepe}, {Mayor}, {Queloz},
  {Bouchy}, {Avila}, {Benz}, {Bertaux}, {Bonfils}, {Dall}, {Delabre}, {Dekker},
  {Eckert}, {Fleury}, {Gilliotte}, {Gojak}, {Guzman}, {Kohler}, {Lizon}, {Lo
  Curto}, {Longinotti}, {Lovis}, {Megevand}, {Pasquini}, {Reyes}, {Sivan},
  {Sosnowska}, {Soto}, {Udry}, {Van Kesteren}, {Weber}, \& {Weilenmann}}]{R04}
{Rupprecht}, G., {Pepe}, F., {Mayor}, M., {et~al.} 2004, in Society of
  Photo-Optical Instrumentation Engineers (SPIE) Conference Series, Vol. 5492,
  Society of Photo-Optical Instrumentation Engineers (SPIE) Conference Series,
  ed. {A.~F.~M.~Moorwood \& M.~Iye}, 148--159

\bibitem[{{Schlaufman}(2010)}]{Sman10}
{Schlaufman}, K.~C. 2010, ApJ, 719, 602

\bibitem[{{Shporer} {et~al.}(2007){Shporer}, {Tamuz}, {Zucker}, \&
  {Mazeh}}]{Shporer07}
{Shporer}, A., {Tamuz}, O., {Zucker}, S., \& {Mazeh}, T. 2007, MNRAS, 376, 1296

\bibitem[{{Simpson} {et~al.}(2010{\natexlab{a}}){Simpson}, {Baliunas}, {Henry},
  \& {Watson}}]{Simpson10rot}
{Simpson}, E.~K., {Baliunas}, S.~L., {Henry}, G.~W., \& {Watson}, C.~A.
  2010{\natexlab{a}}, MNRAS, 408, 1666

\bibitem[{{Simpson} {et~al.}(2010{\natexlab{b}}){Simpson}, {Pollacco},
  {H{\'e}brard}, {Gibson}, {Barros}, {Boisse}, {Bouchy}, {Cameron}, {Miller},
  {Watson}, \& {Keenan}}]{Simpson10w3}
{Simpson}, E.~K., {Pollacco}, D., {H{\'e}brard}, G., {et~al.}
  2010{\natexlab{b}}, MNRAS, 405, 1867

\bibitem[{{Stempels} {et~al.}(2007){Stempels}, {Collier Cameron}, {Hebb},
  {Smalley}, \& {Frandsen}}]{Stempels07}
{Stempels}, H.~C., {Collier Cameron}, A., {Hebb}, L., {Smalley}, B., \&
  {Frandsen}, S. 2007, MNRAS, 379, 773

\bibitem[{{Street} {et~al.}(2010){Street}, {Simpson}, {Barros}, {Pollacco},
  {Joshi}, {Todd}, {Collier Cameron}, {Enoch}, {Parley}, {Stempels}, {Hebb},
  {Triaud}, {Queloz}, \& {Segransan}}]{Street10}
{Street}, R.~A., {Simpson}, E., {Barros}, S.~C.~C., {et~al.} 2010, ApJ, 720,
  337

\bibitem[{{Triaud} {et~al.}(2010){Triaud}, {Collier Cameron}, {Queloz},
  {Anderson}, {Gillon}, {Hebb}, {Hellier}, {Loeillet}, {Maxted}, {Mayor},
  {Pepe}, {Pollacco}, {S{\'e}gransan}, {Smalley}, {Udry}, {West}, \&
  {Wheatley}}]{Triaud10}
{Triaud}, A.~H.~M.~J., {Collier Cameron}, A., {Queloz}, D., {et~al.} 2010,
  ArXiv e-prints

\bibitem[{{Triaud} {et~al.}(2009){Triaud}, {Queloz}, {Bouchy}, {Moutou},
  {Cameron}, \& {et al.}}]{Triaud09}
{Triaud}, A.~H.~M.~J., {Queloz}, D., {Bouchy}, F., {et~al.} 2009, A\&A, 506,
  377

\bibitem[{{Valenti} \& {Fischer}(2005)}]{VF05}
{Valenti}, J.~A. \& {Fischer}, D.~A. 2005, ApJS, 159, 141

\bibitem[{{Watson} {et~al.}(2010{\natexlab{a}}){Watson}, {Littlefair},
  {Cameron}, {Dhillon}, \& {Simpson}}]{Watson10sini}
{Watson}, C.~A., {Littlefair}, S.~P., {Cameron}, A.~C., {Dhillon}, V.~S., \&
  {Simpson}, E.~K. 2010{\natexlab{a}}, MNRAS, 408, 1606

\bibitem[{{Watson} {et~al.}(2010{\natexlab{b}}){Watson}, {Littlefair},
  {Diamond}, {Collier Cameron}, {Fitzsimmons}, {Simpson}, {Moulds}, \&
  {Pollacco}}]{Watson10disk}
{Watson}, C.~A., {Littlefair}, S.~P., {Diamond}, C., {et~al.}
  2010{\natexlab{b}}, ArXiv e-prints

\bibitem[{{Wheatley} {et~al.}(2010){Wheatley}, {Collier Cameron}, {Harrington},
  {Fortney}, {Simpson}, {Anderson}, {Smith}, {Aigrain}, {Clarkson}, {Gillon},
  {Haswell}, {Hebb}, \& {H{\'e}brard}}]{Wheatley10}
{Wheatley}, P.~J., {Collier Cameron}, A., {Harrington}, J., {et~al.} 2010,
  ArXiv e-prints

\bibitem[{{Winn} {et~al.}(2010{\natexlab{a}}){Winn}, {Fabrycky}, {Albrecht}, \&
  {Johnson}}]{Winn10}
{Winn}, J.~N., {Fabrycky}, D., {Albrecht}, S., \& {Johnson}, J.~A.
  2010{\natexlab{a}}, ApJL, 718, L145

\bibitem[{{Winn} {et~al.}(2009){Winn}, {Howard}, {Johnson}, {Marcy}, {Gazak},
  \& {et~al.}}]{Winn09b}
{Winn}, J.~N., {Howard}, A.~W., {Johnson}, J.~A., {et~al.} 2009, ApJ, 703, 2091

\bibitem[{{Winn} {et~al.}(2010{\natexlab{b}}){Winn}, {Howard}, {Johnson},
  {Marcy}, {Isaacson}, {Shporer}, {Bakos}, {Hartman}, {Holman}, \&
  {Albrecht}}]{Winn10hatp14}
{Winn}, J.~N., {Howard}, A.~W., {Johnson}, J.~A., {et~al.} 2010{\natexlab{b}},
  ArXiv e-prints

\bibitem[{{Winn} {et~al.}(2010{\natexlab{c}}){Winn}, {Johnson}, {Howard},
  {Marcy}, {Isaacson}, {Shporer}, {Bakos}, {Hartman}, \&
  {Albrecht}}]{Winn10hatp11}
{Winn}, J.~N., {Johnson}, J.~A., {Howard}, A.~W., {et~al.} 2010{\natexlab{c}},
  ApJL, 723, L223

\bibitem[{{Winn} {et~al.}(2007){Winn}, {Johnson}, {Peek}, {Marcy}, {Bakos},
  {Enya}, {Narita}, {Suto}, {Turner}, \& {Vogt}}]{Winn07}
{Winn}, J.~N., {Johnson}, J.~A., {Peek}, K.~M.~G., {et~al.} 2007, ApJL, 665,
  L167

\bibitem[{{Wu} \& {Murray}(2003)}]{WM03}
{Wu}, Y. \& {Murray}, N. 2003, ApJ, 589, 605

\end{thebibliography}

\end{document}